\documentclass[a4paper,11pt]{article}
\pdfoutput=1 

\usepackage{jheppub} 

\usepackage[T1]{fontenc} 

\usepackage{caption}
\usepackage{color}
\usepackage{graphicx}
\usepackage{amsmath}
\usepackage{amssymb}

\hypersetup{allcolors=[rgb]{0,0,1}}
\newcommand{\orcid}[1]{\href{https://orcid.org/#1}{#1}}
\usepackage{scalerel}
\usepackage{graphicx}
\usepackage[normalem]{ulem}
\usepackage{subcaption}
\usepackage{lineno}
\usepackage{soul}

\newcommand{\olsi}[1]{\,\overline{\!{#1}}} 

\title{\boldmath Dark Photon Search at Yemilab, Korea}

\author[a,1,2]{S. H. Seo,\note{Corresponding author}\note{\orcid{orcid \# 0000-0002-1496-624X}}}
\author[a,b,c,3]{Y. D. Kim\note{\orcid{orcid \# 0000-0003-2471-8044}}}

\affiliation[a]{Center for Underground Physics, Institute for Basic Science (IBS), \\
  55, Expo-ro, Yuseong-gu, Daejeon, 34126, Korea}
\affiliation[b]{Department of Physics and Astronomy, Sejong University, Seoul, 05006, Korea}
\affiliation[c]{IBS School, University of Science and Technology (UST), Daejeon, 34113, Korea}

\emailAdd{sunny.seo@ibs.re.kr}
\emailAdd{ydkim@ibs.re.kr}

\abstract{
  Dark photons are well motivated hypothetical dark sector particles that could account for observations
  that cannot be explained by the standard model of particle physics.  A search for dark photons that are
  produced by an electron beam striking a thick tungsten target and subsequently
  interact in a 3~kiloton-scale neutrino detector in Yemilab, a new underground lab in Korea, is proposed.  
  Dark photons can be produced by ``darkstrahlung'' or by oscillations from ordinary photons produced in
  the  target and detected by their visible decays, ``absorption'' or by their oscillation to ordinary photons.
  By detecting the absorption process or the oscillation-produced photons, a world's best sensitivity
  for measurements of the dark-photon kinetic mixing parameter of
  $\epsilon^2 > 1.5 \times 10^{-13} (6.1 \times 10^{-13})$ at the 95\% confidence level (C.L.) could be obtained
  for dark photon masses between 80~eV and 1~MeV in  a year-long exposure to a 100~MeV-100~kW electron beam
  with zero ($10^3$) background events. 
  In parallel, the detection of $e^+e^-$ pairs from decays of dark photons with mass
  between 1~MeV and $\sim$86~MeV would have sensitivities of 
  $\epsilon^2 > \mathcal{O}(10^{-17}) (\mathcal{O}(10^{-16}))$ at the 95\% C.L.
  with zero ($10^3$) background events.  
  This is comparable to that of the Super-K experiment under the same zero background assumption.
}

\begin{document} 
\maketitle
\flushbottom

\section{Introduction}
\label{sec:intro}

The standard model of particle physics has been very successful at explaining phenomena in the
visible universe.  However, it has a number of clear limitations.  It provides no explanations for
dark matter, the muon g-2 anomaly,  the $m_{ee}=17$~MeV $e^+e^-$ excess from $^8$Be that is reported in
refs.~\cite{Krasznahorkay:2015iga,Krasznahorkay:2019lyl}, etc.  In order to explain these ``beyond
the standard model'' (BSM) hints of new physics, the introduction of a hypothetical sector of new particles
and interactions, the dark (or hidden) sector  has been proposed. In this scheme, there are only a
few portals (or mediators) that connect the dark sector to the visible universe that have significant
strength and satisfy Lorentz and gauge symmetries.  These are the vector, higgs, neutrino, and axion
portals that can be explored with different types of experiments. 

In particular, in the vector portal, a dark photon (usually denoted as $A^{\prime}, \phi$, or $\gamma^{\prime}$),
that usually described by an extra $U(1)_D$ gauge symmetry group, is the dark sector particle  
that could be more readily explored than corresponding particles in the other portals because it can kinetically
mix with an ordinary photon. As a result, any experiment that can produce photons and detect photons or
leptons can, in principle, explore the vector portal~\cite{Bjorken:2009mm}.

The Lagrangian that describes the dark photon (DP) is   
\begin{eqnarray}
  \mathcal{L} \supset - \frac{1}{4}F_{\mu\nu}^{\prime}F^{\prime\mu\nu} + \frac{\epsilon}{2}F_{\mu\nu}^{\prime}F^{\mu\nu}
  + \frac{m_{\phi}^2}{2}A_{\mu}^{\prime}A^{\prime\mu},
\label{e:ldp}
\end{eqnarray}
where $F_{\mu\nu}^{\prime} \equiv \partial_\mu A_\nu^\prime - \partial_\nu A_\mu^\prime$ is the DP field strength tensor,
$A_\mu^\prime$ is the $U(1)_D$ gauge field,  $\epsilon$ is the kinetic-mixing strength between the dark and
ordinary photons, and $m_\phi$ is the dark photon mass. 

\begin{figure*}[t]
\begin{center}
\includegraphics[width=0.95\textwidth]{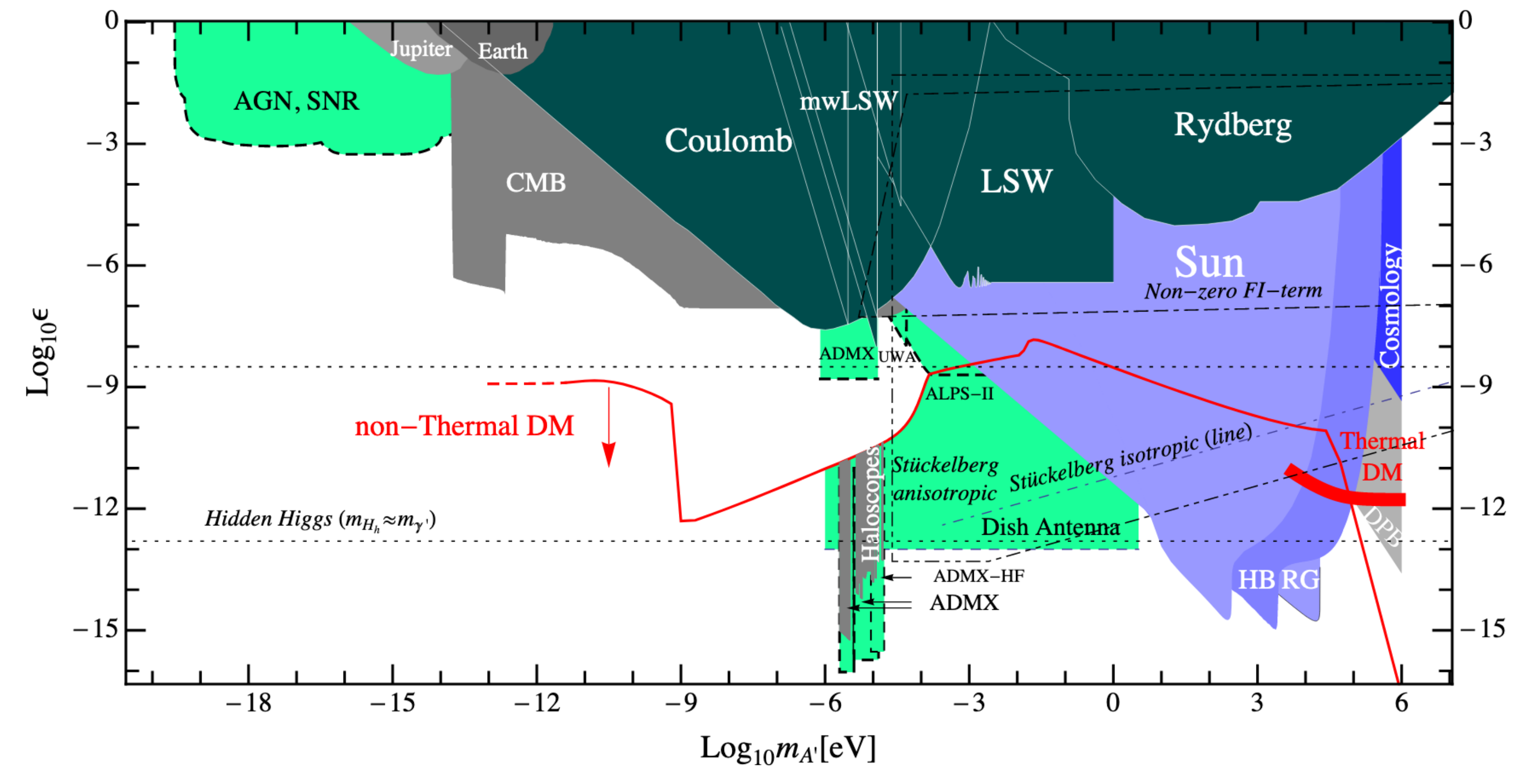}
\end{center}
\caption{
  Current limits on $\epsilon$ for $m_\phi < 2m_e$ dark photons from laboratory search
  experiments and  astrophysical observations (from Ref. \cite{Essig:2013lka}).
}
\label{f:dp_lim} 
\end{figure*}

\begin{figure*}
\begin{center}
\includegraphics[width=0.75\textwidth]{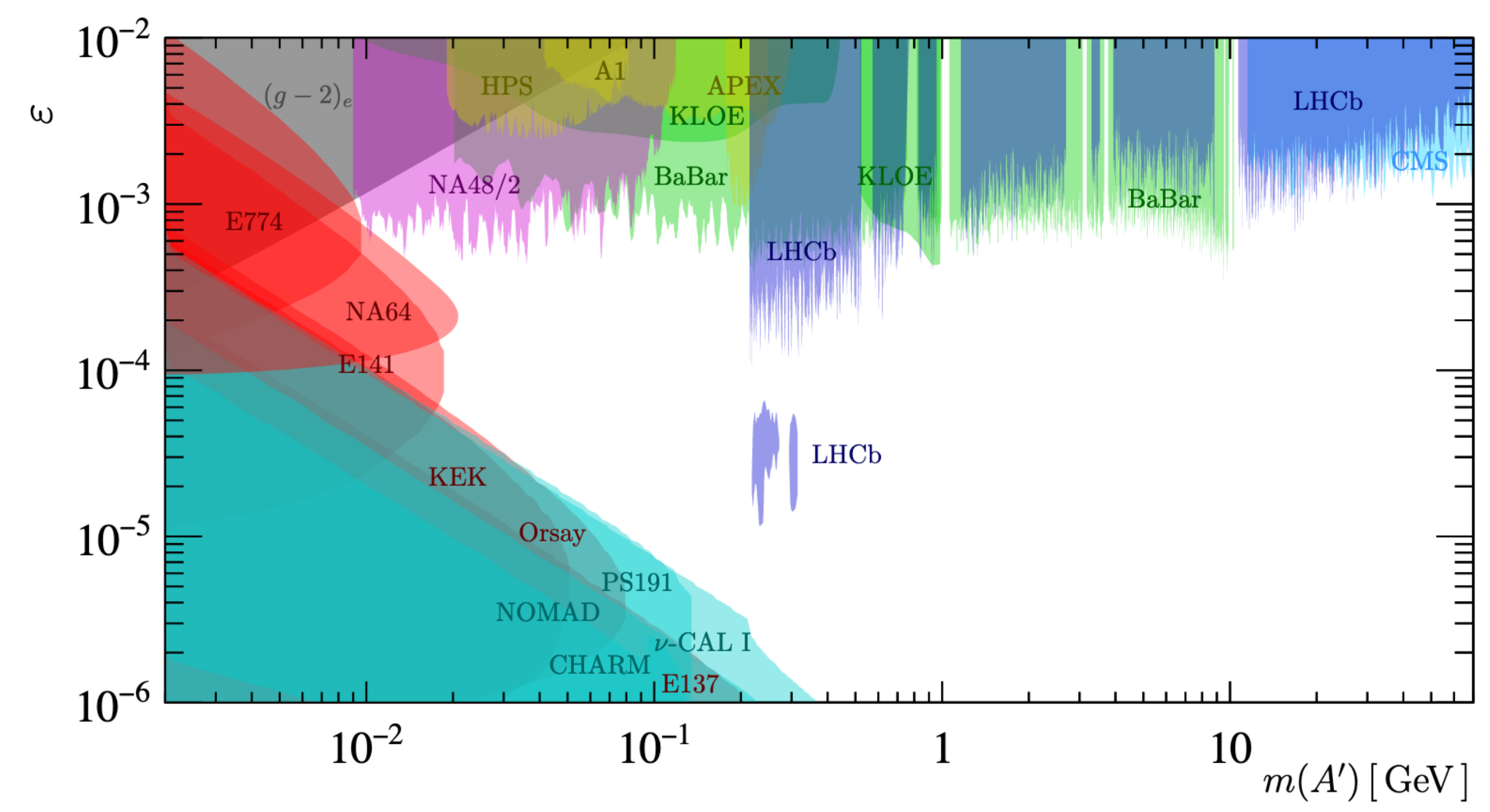}
\includegraphics[width=0.75\textwidth]{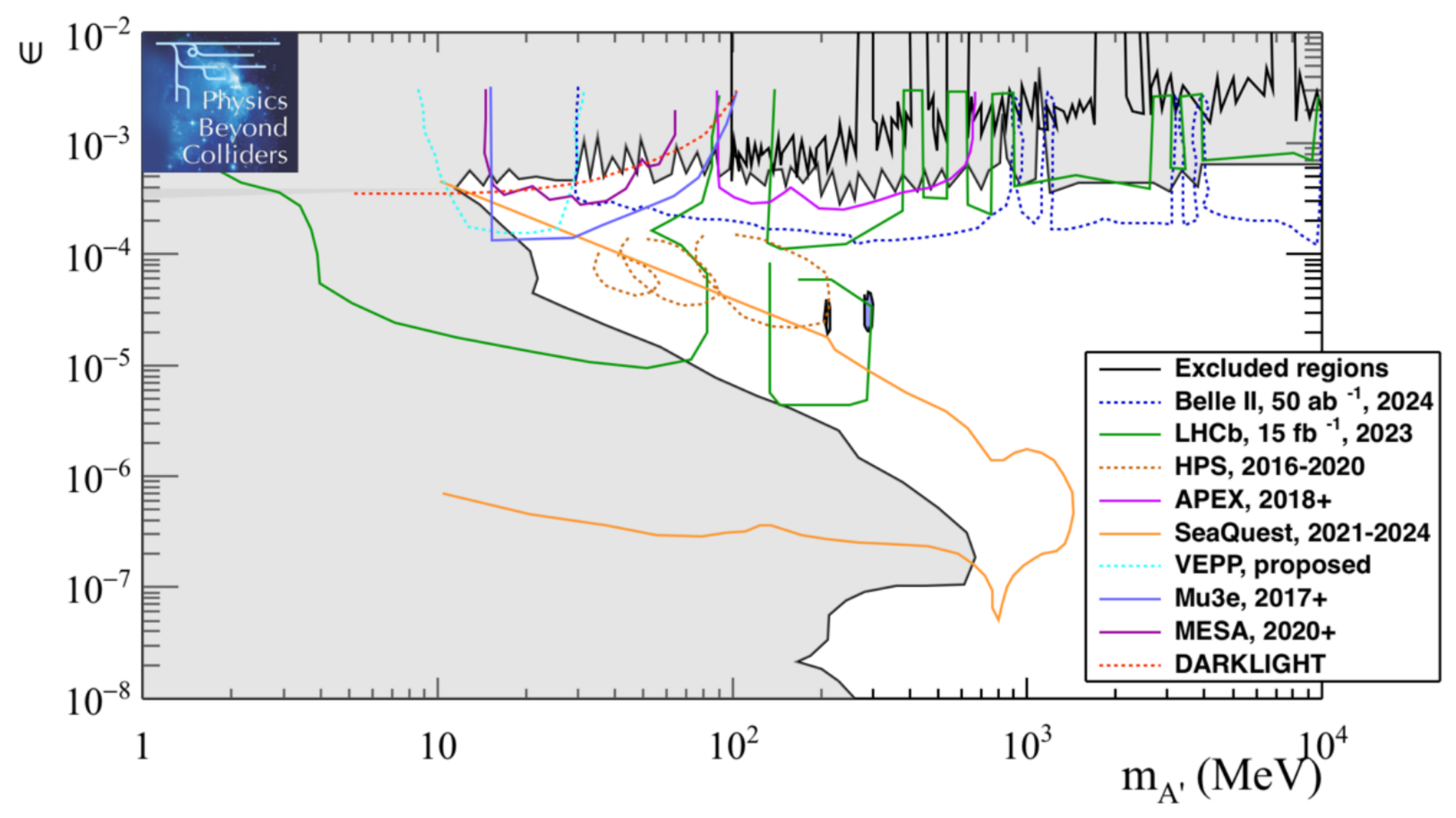}
\includegraphics[width=0.75\textwidth]{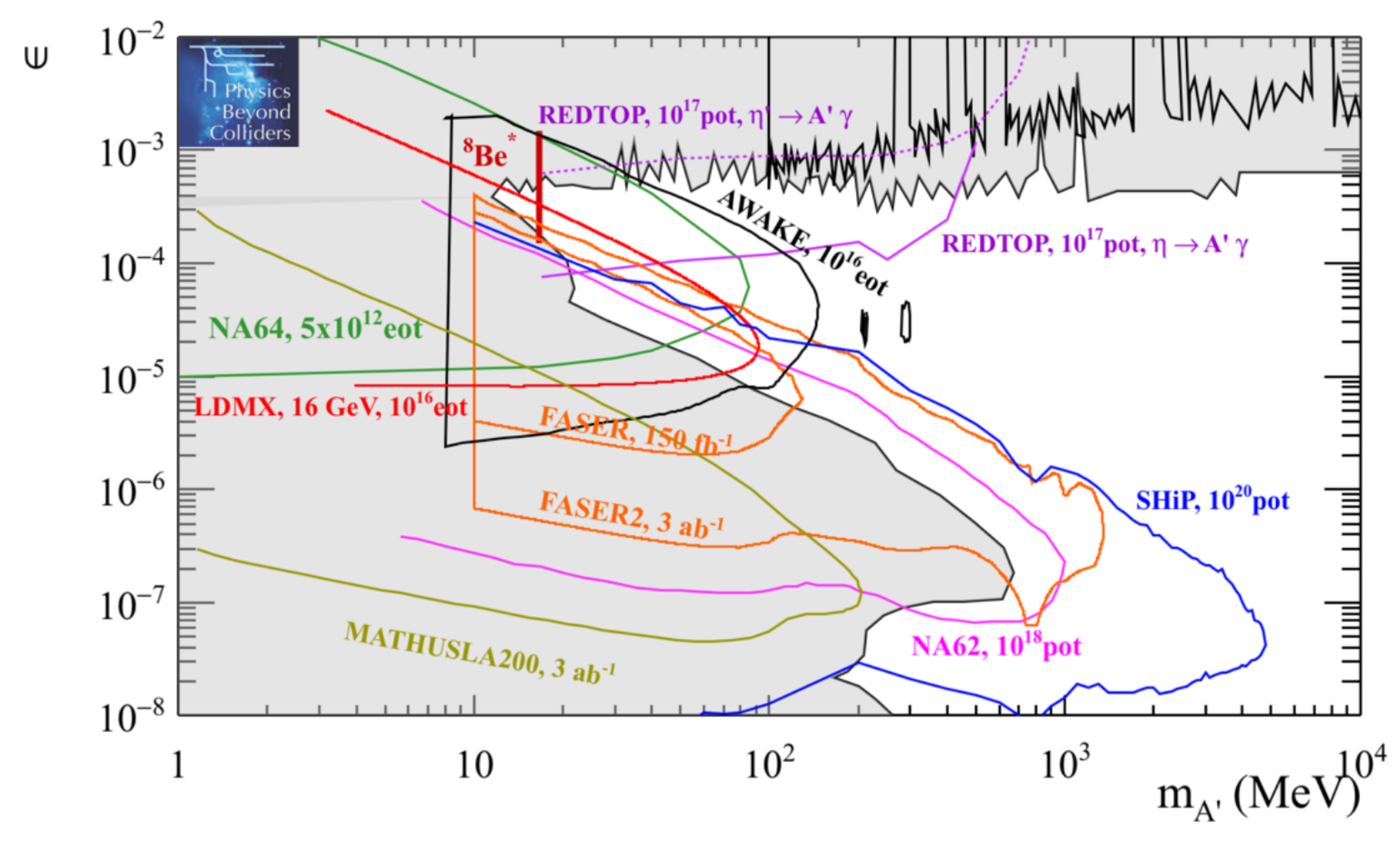}
\end{center}
\caption{
  Current limits (color-filled or shaded regions) and sensitivities (solid lines) on $\epsilon$ for $m_\phi > 2m_e$ dark photons from various
  experiments. Current limits only (top) from Ref. \cite{Aaij:2019bvg}, 
  the sensitivities that are expected to be by the experiments/proposals outside Physics Beyond Colliders (PBC) activities (middle), 
  and those within PBC activities (bottom) from Ref. \cite{Beacham:2019nyx}. Up-to-date limits can be also directly obtained in a software framework 
  called darkcast \cite{Ilten:2018crw} at \href{https://gitlab.com/philten/darkcast}{https://gitlab.com/philten/darkcast}.
}
\label{f:dp_lim2}
\end{figure*}

The 1988 SLAC beam dump experiment (E137)  was the pioneering search for dark photons~\cite{Bjorken:1988as}. 
More recently, in the last decade, there have been numerous reports of dark photon searches~\cite{Bross:1989mp,Merkel:2011ze,Abrahamyan:2011gv,Essig:2010xa,Babusci:2012cr,Izaguirre:2013uxa,Babusci:2014sta,Batell:2014mga,Merkel:2014avp,Blumlein:2013cua,Batley:2015lha,Anastasi:2015qla,Ilten:2015hya,Anastasi:2016ktq,Ilten:2016tkc,Battaglieri:2017qen,Lees:2017lec,Corliss:2017tms,Aaij:2017rft,Ilten:2018crw,Ariga:2018uku,Banerjee:2019hmi,Demidov:2018odn,Park:2017prx,Danilov:2018bks}
based on data from fixed target accelerator experiments, $e^+e^-$ colliders, reactors and astrophysical measurements,
that have set stringent limits to the dark photon parameter space. These limits could be further improved or,
possibly, a dark-photon signal could be discovered, by current or future experiments without huge costs. 
Figures~\ref{f:dp_lim} and \ref{f:dp_lim2} show the current constraints (or sensitivities) on
$\epsilon$ for $m_\phi < 2m_e$ and $m_\phi > 2m_e$, respectively; comprehensive reviews on the status of dark
photon searches can be found in refs.~\cite{Beacham:2019nyx,Filippi:2020kii}. 

Yemilab, a new underground lab that is being constructed in Handuk iron mine in Jeongseon-gun, Korea, will have a
cavern that will be capable of hosting a $\sim$3 kiloton liquid target neutrino detector. 
A 100~MeV electron accelerator (100~kW or 10~kW beam power) located close to the neutrino detector, 
would make a dark photon search possible at Yemilab. 

In the following sections, the proposed Yemilab neutrino detector \ref{sec:det}, expected numbers of produced
and detected dark photons \ref{sec:num}, and the expected dark photon sensitivity  \ref{sec:sens} are described.
These are followed by a summary \ref{sec:sum}.

\section{A neutrino detector for Yemilab}
\label{sec:det}

\begin{figure*}[t]
\begin{center}
\includegraphics[width=0.48\textwidth]{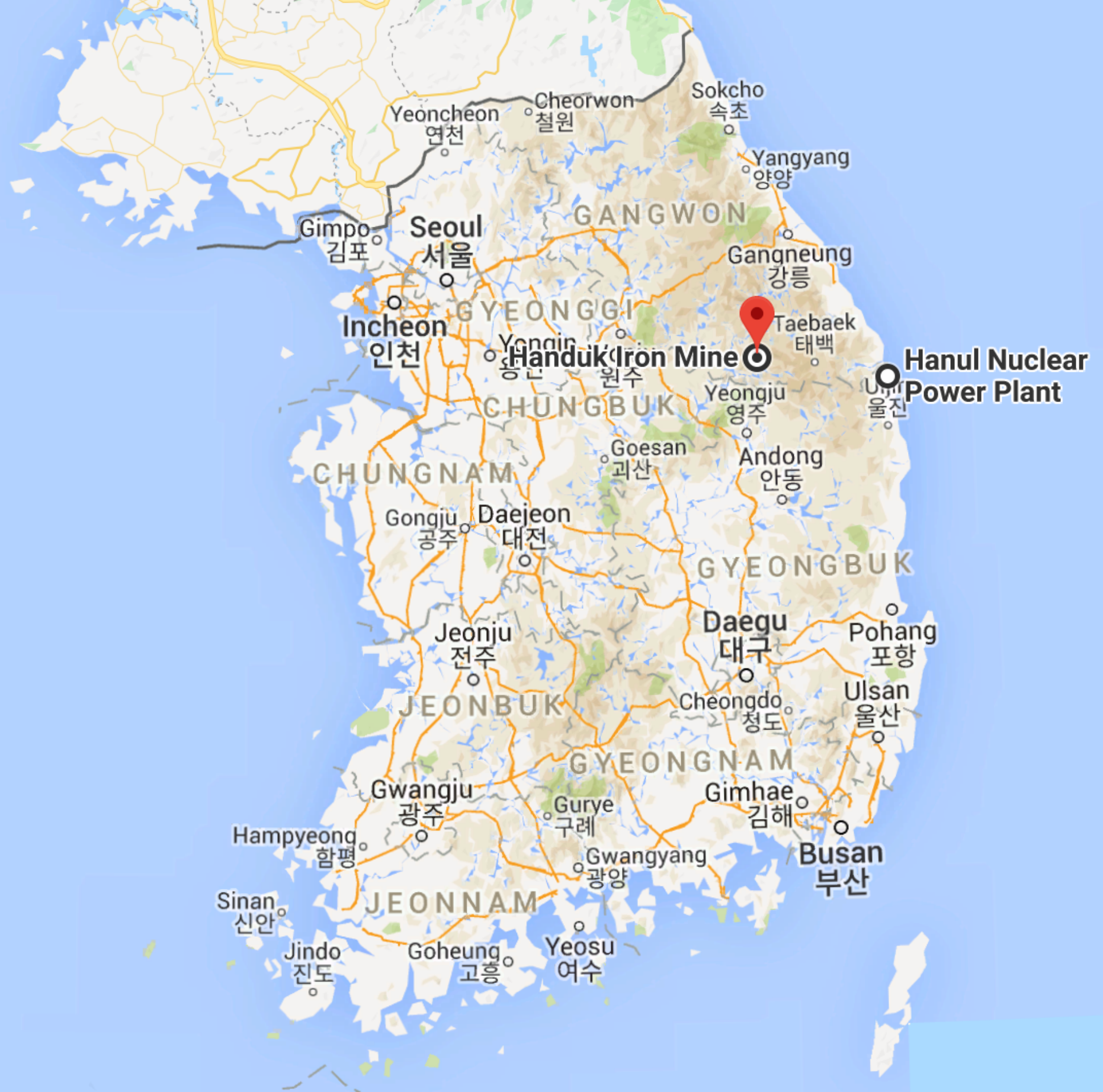}
\includegraphics[width=0.48\textwidth]{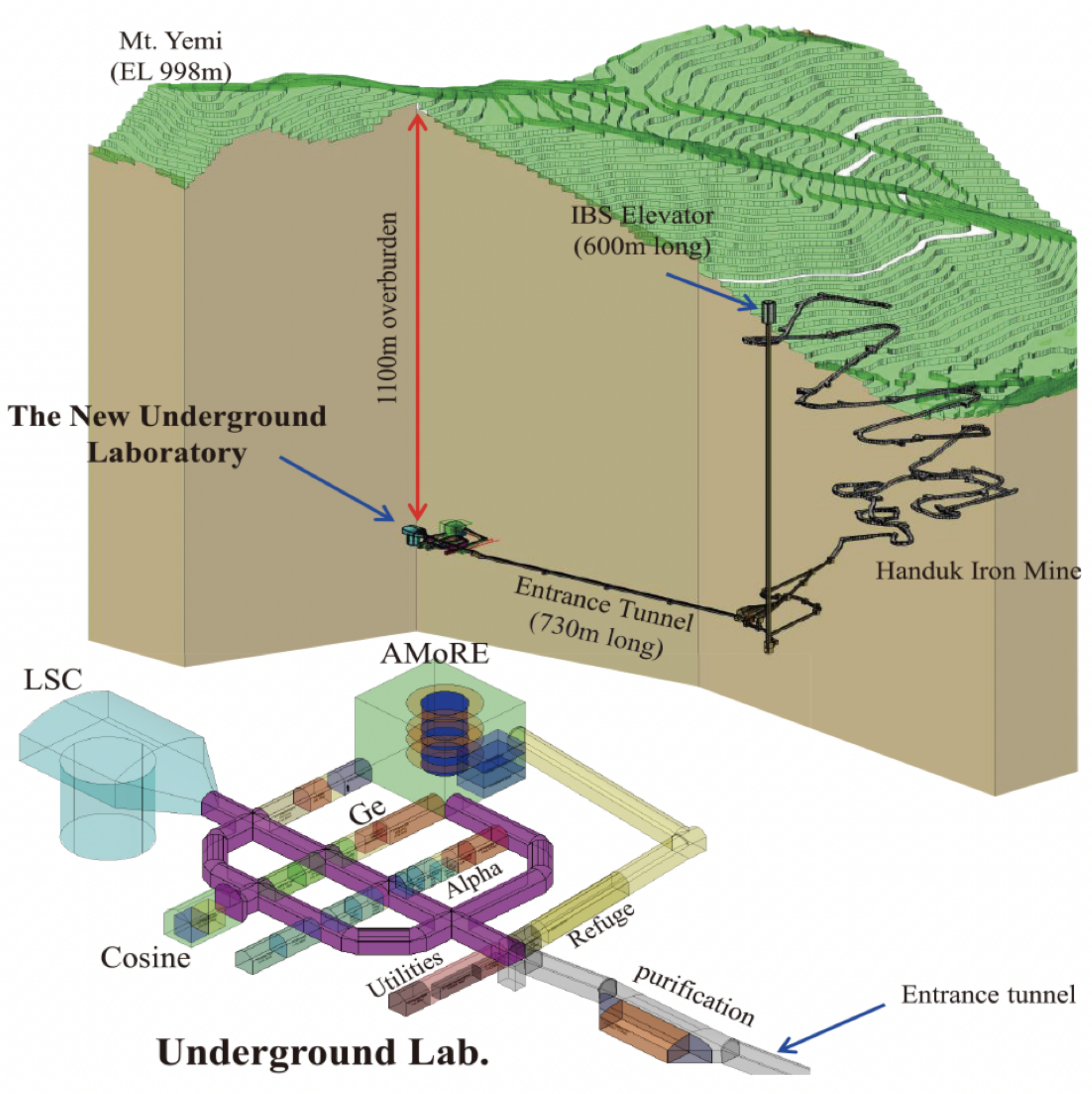}
\end{center}
\caption{
  (Left) The Handuk iron mine location (latitude: 37.188639 deg, longitude: 128.659406 deg) in Jeongseon-gun,
  Gangwon province, Korea, where a new underground Yemilab ($\sim$1000~m overburden) is currently being constructed. 
  (Right) The layout of Yemilab, including a cavern for a $\sim$3 kiloton neutrino detector (LSC). The 
  laboratory will be accessed by a 600~m vertical shaft and a 730~m entrance tunnel.
  Adapted from Ref. \cite{Seo:2019dpr}. 
}
\label{f:yemilab}
\end{figure*}

By early 2022, the 2$^{\rm{nd}}$ phase construction of Yemilab ($\sim$1000~m overburden under the
eponymous Mt. Yemi) will be completed and experimental operation will commence (see Fig.~\ref{f:yemilab}). 
In addition to spaces for the upgraded COSINE~\cite{Adhikari:2019off} dark matter and
AMoRE-II~\cite{Kim:2017xrs} $0\nu\beta\beta$ search experiments, a cavern suitable for hosting a
$\sim$3~kiloton neutrino detector will be available. The current plans for this space include
a liquid scintillator (LS) or water-based LS (WbLS) neutrino detector~\cite{Seo:2019dpr}.

\begin{figure*}[t]
\begin{center}
\includegraphics[width=0.85\textwidth]{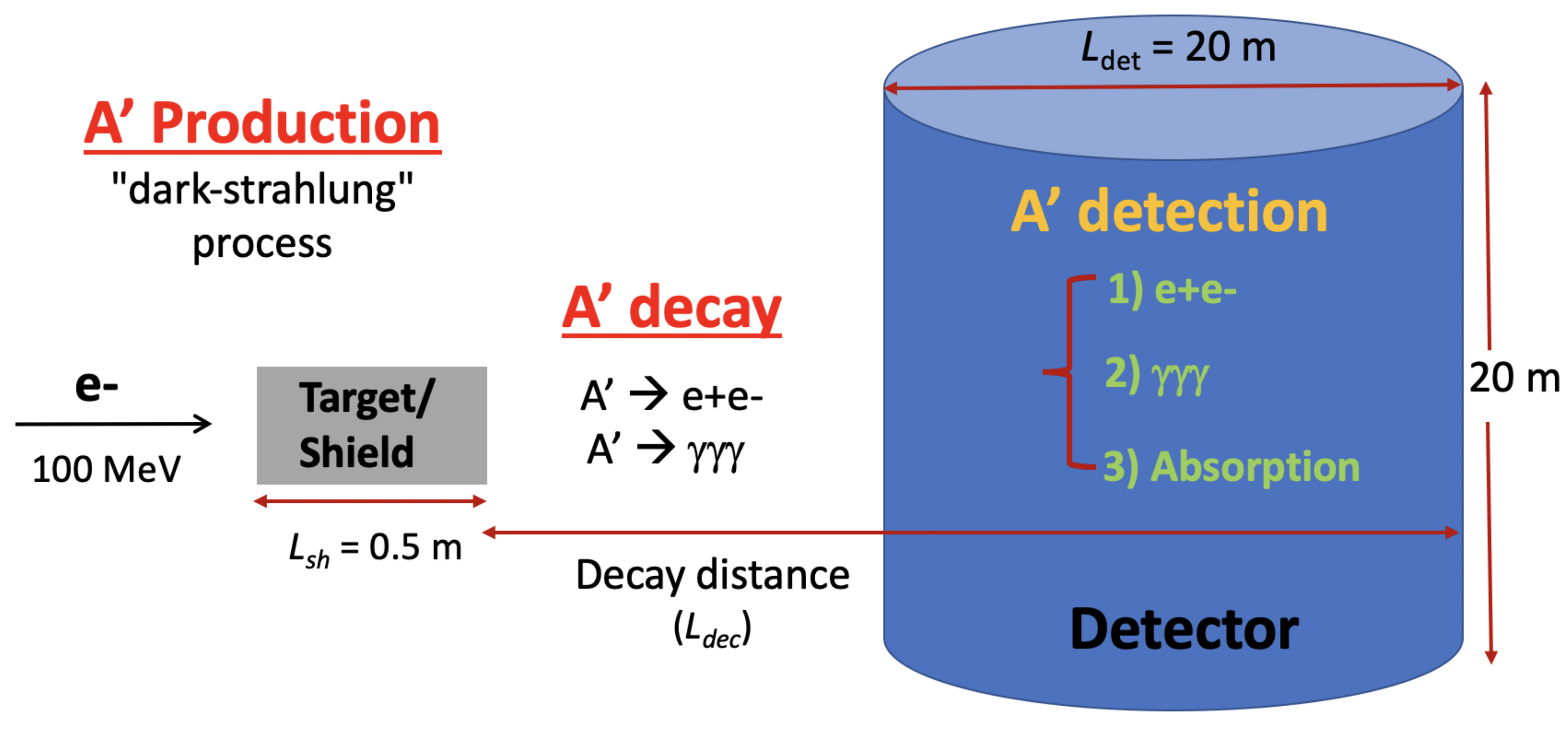}
\end{center}
\caption{
A schematic diagram showing a possible experimental configuration for a dark photon search at Yemilab. 
The $L_{sh}$ ($L_{det}$) represents horizontal length of target \& shield (detector) in the direction of $e^-$ beam. 
Decay distance, $L_{dec}$, is set as 20~m in our study but it is flexible to change. 
}
\label{f:detector}
\end{figure*}

The $\sim$3~kiloton Yemilab neutrino detector is primarily dedicated to precise determinations of solar neutrino
fluxes and measurements of geo-neutrinos~\cite{Seo:2019dpr} 
as well as a sterile neutrino search using a strong radioactive source in a later stage. 
It would be the first kiloton-scale neutrino
telescope in Korea and a follow-on to the successful programs of the smaller scale RENO~\cite{Seo:2016uom} and
NEOS~\cite{Ko:2016owz} reactor neutrino experiments at the Hanbit Nuclear Power Plant. 
The Yemilab neutrino detector could also be used to for a dark photon search by the addition of an electron
accelerator for underground experiments, as suggested in Ref.~\cite{Izaguirre:2015pva}. 
Figure~\ref{f:detector} is a schematic diagram that shows how an electron linac (100~MeV, 100~kW or 10~kW),
tungsten target \& radiation shield (50~cm-thick), 
and  the neutrino detector (D: 20~m, H: 20~m cylinder) could be configured at Yemilab.
To prevent dark photon decay particles ($e^+e^-$ or 3$\gamma$) occurred outside detector target from being absorbed by the outer shell of the detector
or outer detector (i.e., water Cherenkov detector for muon veto and buffer region where PMTs are mounted),
a guiding tube penetrating the outer detector up to the target would be installed.

Final state of dark photon signal would be $e^+e^-$ or single (triple) gamma(s) and this is described in the following section. 
Regardless of the choice of the neutrino target medium between LS and WbLS, our search strategy is inclusive for all possible final states 
for a visible mode considered in this study, and signal candidates are identified by counting number of PMTs fired 
or measuring visible energy within pulsed beam time window. 
Background events are subtracted by beam-ON minus beam-OFF data.

%
\section{Dark Photon Production, Detection and Expected Number}
\label{sec:num}

\subsection{Dark photon production}

\begin{figure*}[t] 
\centering 
\includegraphics[width=0.35\textwidth]{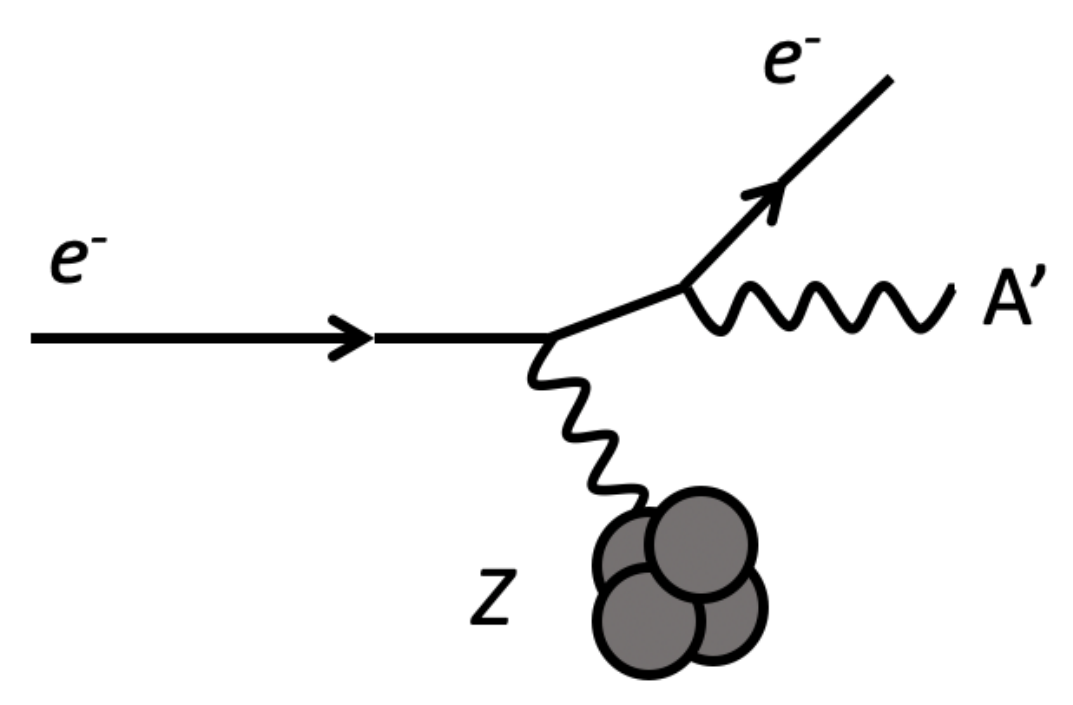}
\hspace{0.7cm}
\includegraphics[width=0.35\textwidth]{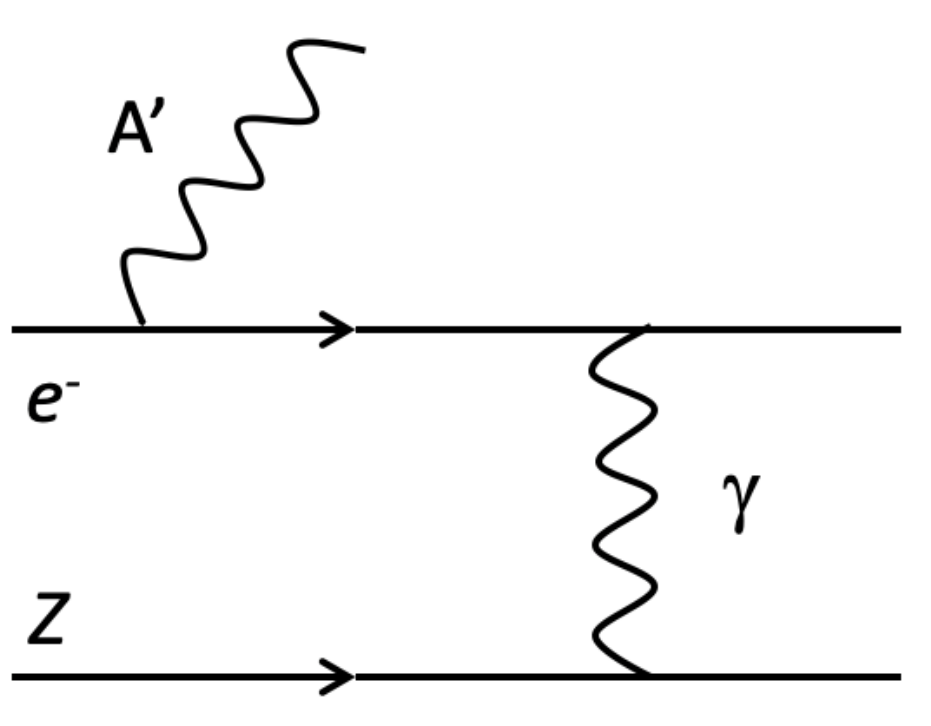}
\caption{\label{f:dp_prod} Dark Photon production via the ``darkstrahlung'' process.
}
\end{figure*}

In dark sector models with a vector portal there is a small mixing between dark photons
and ordinary photons. As a result, dark photons could be produced via the electron
bremsstrahlung process that is common for ordinary photons; i.e., when an electron beam strikes a
target with atomic number $Z$, dark photons, $A^{\prime}$, are produced by the reaction
$e^- + Z \rightarrow e^- + Z + A^{\prime}$, as illustrated in Fig.~\ref{f:dp_prod}.  Practical calculations
generally use the Weiz\"{a}scker-Williams (WW) approximation to obtain the dark photon production cross-section
via this reaction, in this case known as the ``darkstrahlung'' process, as a simplification of the exact result,
which is quite complicated.  For computer calculations an improved WW (IWW) approximation that increases the
calculation speed has been also developed and widely used as well. These approximations, however, assume that the dark photon's mass
is much greater than electron mass and much less than electron beam energy.  Recently, Liu and
Miller~\cite{Liu:2017htz} have reported an exact calculation of the darkstrahlung cross-section and, in 
addition, generalized versions of the WW and IWW approximations so that no restrictions on the dark photon
mass apply.  Liu and Miller compared their exact calculation with the generalized IWW approximation that was
used by SLAC experiment E137 and found reasonable agreement for dark photon masses below 100~MeV.

In this work, the generalized IWW approximation by Liu and Miller is employed to
compute darkstrahlung cross-sections.  This has the following analytic form  

\begin{eqnarray}
  \frac{d\sigma}{dx} = 2\epsilon^2\alpha^3\chi\frac{|k|}{E}
  \frac{m_e^2x(-2+2x+x^2)-2(3-3x+x^2)x\tilde{u}_{max}}{3x\tilde{u}_{max}}, 
\label{e:prod-xsec}
\end{eqnarray} 

\noindent
where $x$ is the fraction of energy a dark photon carries away from electron energy ($E$),
$|k|$ is the 3-momentum of the dark photon, $\epsilon$ is the kinetic mixing parameter, $\alpha (\simeq 1/137)$
is the fine structure constant, 
$\tilde{u}_{max} = -m_{\phi}^2\frac{1-x}{x} - m_e^2x$, 
and $\chi$ is the effective photon flux and given by 

\begin{eqnarray}
\chi = \int_{t_{min}}^{t_{max}}dt \frac{t-t_{min}}{t^2}[G_{2,el}(t)+G_{2,in}(t)]. 
\label{e:dp_flux}
\end{eqnarray}
Here $t$ is the square of the four-momentum transferred to the target nucleus, which ranges from
$t_{min} = \left(\frac{m_\phi^2}{2E} \right)^2$ to $t_{max} = m_\phi^2 + m_e^2$ in the IWW approximation
where the produced dark photon is assumed to be collinear to the incident electron,
and $G_{2,el}(t)$ and $G_{2,in}(t)$ are elastic and inelastic form factors of the target nucleus, respectively. 
In the following, we only consider the elastic form factor because the contribution from the inelastic
one is negligibly small.  The $\chi/Z^2$ values are shown in Fig.~10 of Ref.~\cite{Bjorken:2009mm}
for 200~MeV, 1~GeV, and 6~GeV $e^-$ beams on a tungsten target.  For $m_{\phi} < $ 100~MeV with a 200~MeV
$e^-$ beam, the $\chi/Z^2$ values range from $\sim$1 to $\sim$7  and we infer from the figure that the range of
$\chi/Z^2$ values for MeV-scale dark photon masses are similar for the case of a 100~MeV $e^-$ beam. 
For simplicity, in this work we use $\chi/Z^2 = 6$ and found that changes in this value over a
reasonable range has negligible effects on our results. 

\begin{figure*}[t]
\centering
\includegraphics[width=0.93\textwidth]{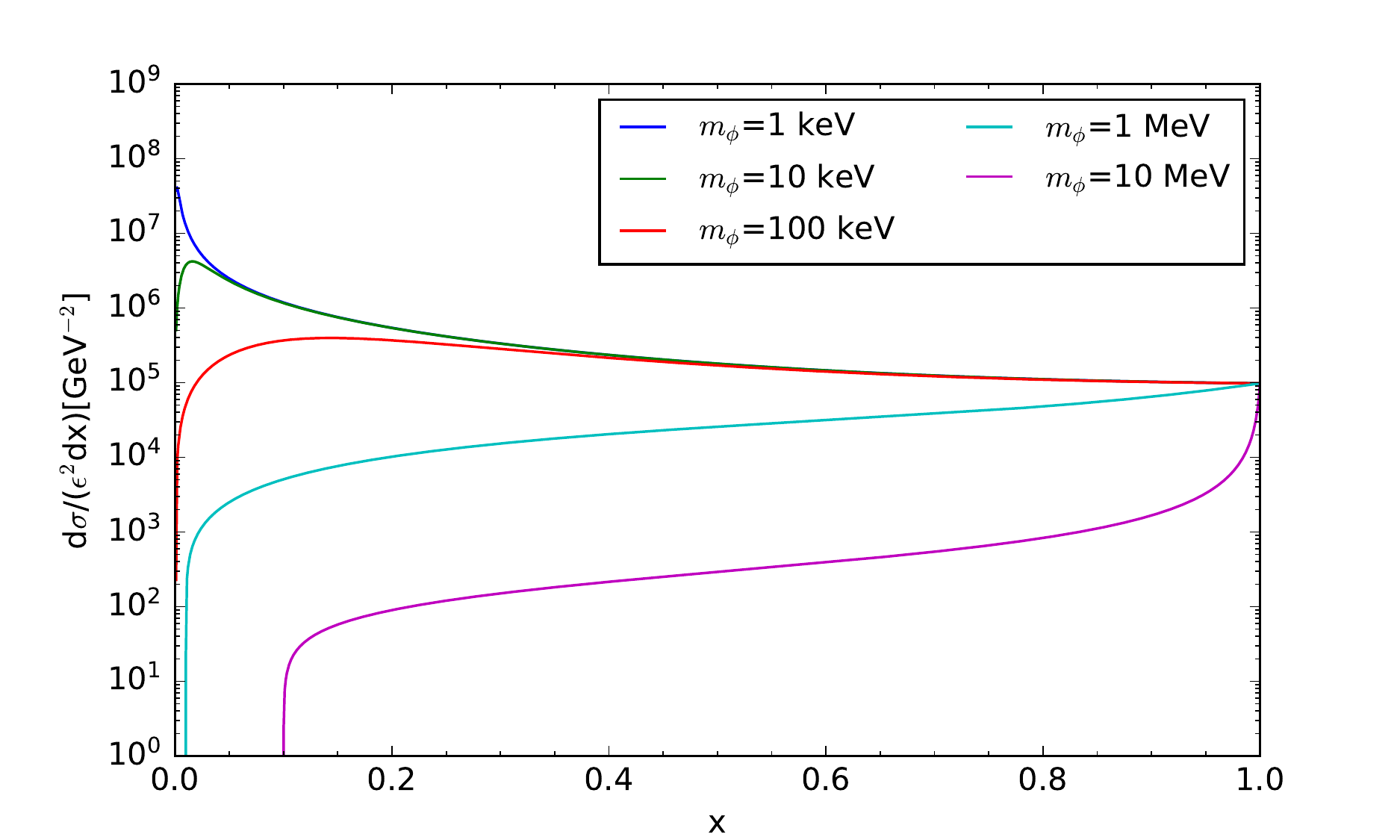}
\caption{\label{f:dp_xsec}
  The generalized IWW differential cross-sections for DP production for a 100~MeV $e^-$ beam on a tungsten target
  for several different DP masses.  Here $x$ is the fraction of DP energy relative to that of the e$^{-}$ beam energy. 
  Note that here the differential cross-section is scaled by $1/\epsilon^2$. 
}
\end{figure*}

Figure~\ref{f:dp_xsec} shows the differential darkstrahlung cross-section for a 100~MeV $e^-$ beam
on a tungsten target as a function of the fractional dark photon (DP) energy, $x$, 
where the cross-sections of the DP masses of 1~keV, 10~keV, 100~keV, 1~MeV, and 10~MeV are compared. 
As the dark photon mass gets heavier, the cross-section decreases as expected from Eq.~(\ref{e:prod-xsec}),
while the relative contribution from high $x$ values increases. 

\subsection{Dark photon detection}

If  $m_{\phi} > 2m_e$, dark photons could decay to the visible $e^+e^-$ final state; if $m_{\phi} < 2m_e$
the only visible decay mode is the $\gamma\gamma\gamma$ channel. If the DP mass is greater than $2m_{\mu}$,
it could also decay to $\mu^+\mu^-$. 
In this study, the e$^-$ beam energy is taken to be 100~MeV and only the $e^+e^-$
and $3\gamma$ decay modes are considered. The expected DP decay length in the lab frame is \cite{Liu:2017htz} 

\begin{eqnarray}
l_{\phi} = \frac{E_k}{m_{\phi}} \frac{1}{\Gamma_{\phi}}, 
\label{e:decay_length}
\end{eqnarray}

\noindent
where $E_k$ is dark photon energy, and $\Gamma_{\phi}$ is the decay width of a dark photon and given in Eqs.~(\ref{e:width_ee}) and (\ref{e:width_3g}) 
for $e^+e^-$ and $\gamma\gamma\gamma$ decays, respectively: 

\begin{eqnarray}
  \Gamma (\phi \rightarrow e^+e^-) = \epsilon^2\frac{\alpha}{2}m_{\phi}\left(1+\frac{2m_e^2}{m_{\phi}^2}
  \right)\left(1-\frac{4m_e^2}{m_\phi^2} \right)^{1/2},
\label{e:width_ee}
\end{eqnarray}

\begin{eqnarray}
  \Gamma (\phi \rightarrow \gamma\gamma\gamma) =
  \epsilon^2\frac{\alpha^4}{2^73^65^2\pi^3}\frac{m_\phi^9}{m_e^8}\left[\frac{17}{5}
    +\frac{67}{42}\frac{m_\phi^2}{m_e^2}+\frac{128941}{246960}\frac{m_\phi^4}{m_e^4}
    + \mathcal{O}\left(\frac{m_\phi^6}{m_e^6} \right) \right].
\label{e:width_3g}
\end{eqnarray}

\begin{figure}[!tbp]
\centering
\includegraphics[width=0.6\textwidth]{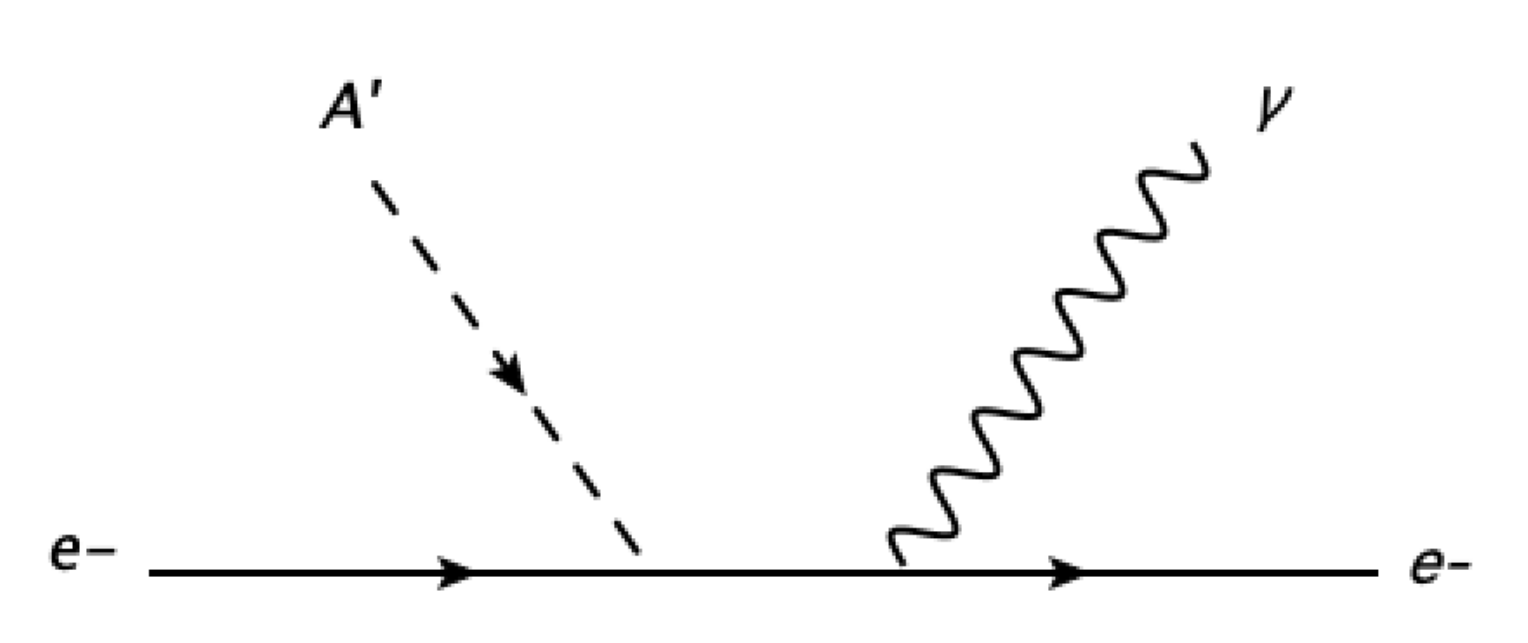}
\caption{\label{f:dp_det} Dark photon absorption process similar to Compton scattering}
\end{figure}

Dark photons can also interact with electrons in the material of the target, shield, and detector, thereby producing
real photons in a process similar to Compton scattering as shown in Fig.~\ref{f:dp_det}; this is
called dark photon ``absorption.''  The DP absorption length in the lab frame is given by   

\begin{eqnarray}
\lambda = \frac{1}{n_e\sigma_{abs}}, 
\label{e:dp_abs}
\end{eqnarray}

\noindent 
where $n_e$ is the electron number density of the medium and $\sigma_{abs}$ is the total cross-section of the
DP absorption and can be computed using Eq. (36) in Ref.~\cite{Liu:2017htz}.

\subsection{Expected number of dark photons}

Using the DP production cross-section for the darkstrahlung process in a thick target, and detection
through visible decays and absorption interactions that are discussed above, the expected number of dark
photons that are either absorbed or decay in the  detector is given by

\begin{equation}
\begin{aligned}
N_{\phi} \approx & \frac{N_{e}X}{M}\int_{E_{min}}^{E_0}dE \int_{x_{min}}^{x_{max}}dx 
\int_{0}^{T}dt I_e(E_0,E,t)\frac{d\sigma}{dx} \\
 & \times e^{-L_{sh}(\frac{1}{l_{\phi}}+\frac{1}{\lambda_{sh}})}
   \left(1-e^{-(L_{dec}/l_{\phi} + L_{det}/\lambda_{det})}\right), 
\end{aligned}
\label{e:dp_num}
\end{equation}

\noindent 
where $N_e$ is the total number of incoming electrons, $X$  the radiation length of the target material
(6.8~gm/cm$^2$ for tungsten);  $M$ is the mass of target atom; $E_0$ is the incoming electron beam energy;
$E_{\textrm{min}} = m_e + \textrm{max}(m_\phi, E_{cut})$,  $x_{\textrm{min}} = \frac{\textrm{max}(m_\phi, E_{cut})}{E}$,
where $E_{\textrm{cut}}$ is the measured energy cutoff depending on the detector; $x_{\textrm{max}}$ is very close
to, but smaller than, 1 and is approximated to be $1 - \frac{m_e}{E}$ if the DP and the initial and final 
electron states are collinear; $T = \rho L_{sh}/X$ where $\rho$ is the density of the target; $l_\phi$ is the
decay length of the DP in the lab frame; $\lambda_{\textrm{sh}} (\lambda_{\textrm{det}})$ is the absorption length
of the DP passing through target and shield (detector). Even though electrons enter the target with initial energy
$E_0$, DP production could occur after some energy loss of the incoming electrons as they penetrate the target. 
This is taken into account with an analytic function $I_e(E_0,E,t)$ from Ref.~\cite{Liu:2017htz} that was
derived in \cite{Tsai:1966js}:

\begin{equation}
I_e(E_0,E,t) = \frac{\left(ln\frac{E_0}{E} \right)^{\frac{4}{3}t-1}}{E_0\Gamma(\frac{4}{3}t)},
\label{e:e-spec}
\end{equation}

\noindent
where $t$ represents how many numbers of radiation length traversed
by the electron before ``darkstrahlung'' occurs, $E$ is the $e^{-}$ energy after $t$ radiation lengths and
$\Gamma$ is the Gamma function.  For the Yemilab neutrino detector, $E_{0} = 100$~MeV, $E_{\textrm{cut}} = 200$~keV,
$N_e = 1.97 \times 10^{23}$  for 1 year of operation with 100~kW $e^-$ beam power,
$L_{sh} = 50$~cm, $L_{\textrm{dec}} = L_{\textrm{det}} = 20$~m. (The decay distance, $L_{\textrm{dec}}$, could be larger,
depending on the distance between the shield and the detector, but the final sensitivity result does not
change by much when $L_{\textrm{dec}}$ is doubled.)

Note also that DP signal loss can occur because of DP decay or absorption in the shield that is
required to attenuate all of the standard model particle background; this loss in shield is
accounted for by the $e^{-L_{sh}(\frac{1}{l_{\phi}}+\frac{1}{\lambda_{sh}})}$ term in Eq.~(\ref{e:dp_num}). The
DP detection probability via either decay or absorption is accounted for in the last term,
the large brackets in Eq.~(\ref{e:dp_num}).

\section{Dark Photon Sensitivity at Yemilab}
\label{sec:sens}

Before obtaining the DP sensitivity at Yemilab, the DP decay and absorption lengths are compared to check
which process is dominant for different DP masses and kinetic mixing parameters. 
Then rough estimation of background is discussed followed by DP sensitivities for decay-only, absorption-only, and both combined 
cases. 
For light DP ($m_{\phi} < 2m_e$), oscillation between ordinary and dark photons is additionally discussed and
the corresponding sensitivity is also obtained. 

\subsection{Dark photon decay and absorption lengths}

Table~\ref{t:dp_lengths} lists the DP decay and absorption lengths for $\epsilon = 10^{-3}, 10^{-5}$ and $10^{-8}$
cases for 0.1~MeV and 10~MeV DP masses, which are taken as representative for the $m_{\phi}<2m_e$ and
$m_{\phi}>2m_e$ cases corresponding to the 3$\gamma$ and $e^+e^-$ decay modes, respectively.  Note that the
3$\gamma$ decay lengths are very large while the $e^+e^-$ decay lengths are more compatible with the detector
size.  However, for large values of $\epsilon$ (e.g., $\epsilon = 10^{-3}$) the $e^+e^-$ decay length becomes
quite small ($\sim10^{-5}$~m), and DP decays occur primarily in the shield, and well before they reach the detector. 
The absorption lengths in target/shield (both tungsten) and detector material (water, WbLS or LS)
are similar but the length at the target/shield is an order of magnitude shorter due to its higher density. 
Note that the absorption length-scales are shorter than the 3$\gamma$ decay lengths and larger than those
for the $e^+e^-$ decays. 

\begin{table}[!tbp]
\centering
\begin{tabular}{|c|c|c|c|c|}
\hline
Kinetic mixing   & DP mass & Decay length & Absorption length  & Absorption length \\
parameter $\epsilon$  & $m_\phi$ (MeV) & $l_\phi$ (m) & in shield: $\lambda_{sh}$ (m) & in target: $\lambda_{det}$ (m) \\
\hline
$10^{-3}$  & 0.1  & $\sim 10^{18}$  (3$\gamma$)  & $\sim 10^{5}$   & $\sim 10^{6}$   \\
$10^{-3}$  & 10   &  $\sim 10^{-5}$  ($e^+e^-$) &  $\sim 10^{5}$  &  $\sim 10^{6}$  \\
\hline
$10^{-5}$  & 0.1  & $\sim 10^{22}$  (3$\gamma$)  & $\sim 10^{9}$  & $\sim 10^{10}$   \\
$10^{-5}$  & 10   & $< 1$ ($e^+e^-$)  & $\sim 10^{9}$  &  $\sim 10^{10}$  \\
\hline
$10^{-8}$  & 0.1  & $\sim 10^{28}$  (3$\gamma$)  & $\sim 10^{15}$  & $\sim 10^{16}$  \\
$10^{-8}$  & 10   & $\sim 10^{5}$  ($e^+e^-$)  & $\sim 10^{15}$  & $\sim 10^{16}$  \\
\hline
\end{tabular}
\captionof{table}{\label{t:dp_lengths}
  Dark photon decay and absorption length scales for three different kinetic mixing parameters and for two
  different DP  mass values.
  These length scales depend on the fractional of DP energy (x) and electron energy (E) at the DP point
  of production, while for the allowed ranges of x and E  the length scale values given in the table
  do not change. 
}
\end{table}

\begin{figure}[!tbp]
\centering 
\includegraphics[width=.7\textwidth]{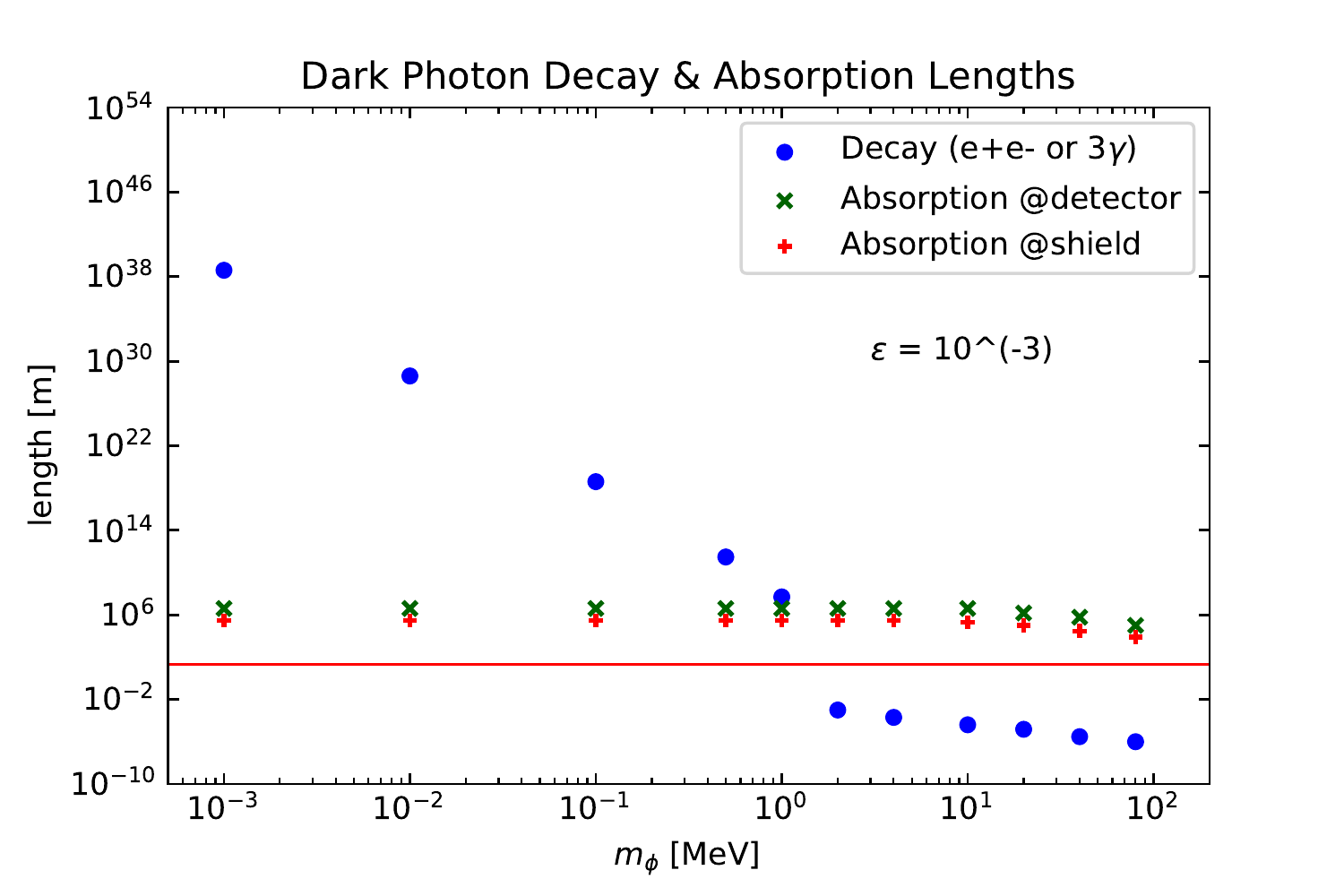}
\includegraphics[width=.7\textwidth]{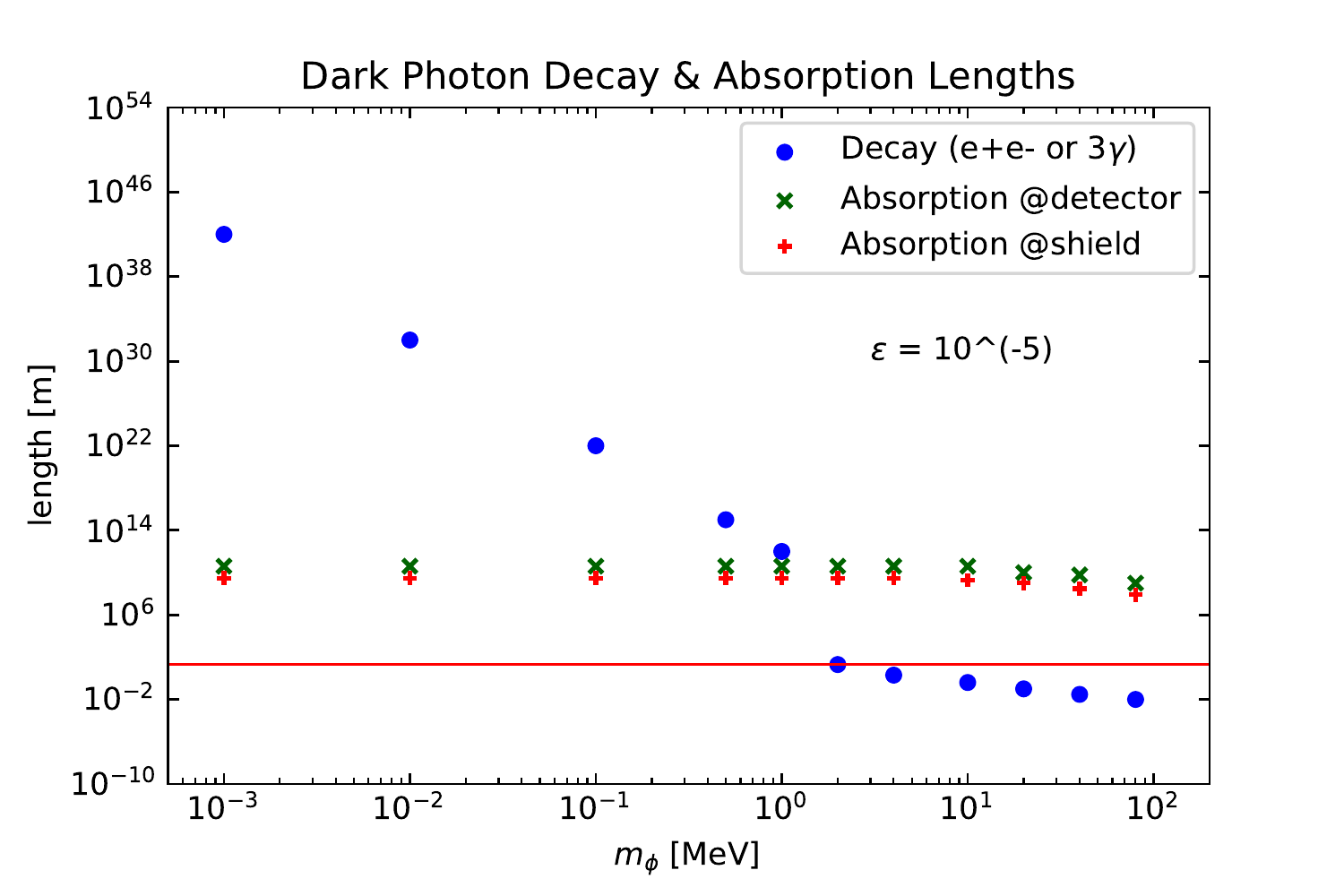}
\includegraphics[width=.7\textwidth]{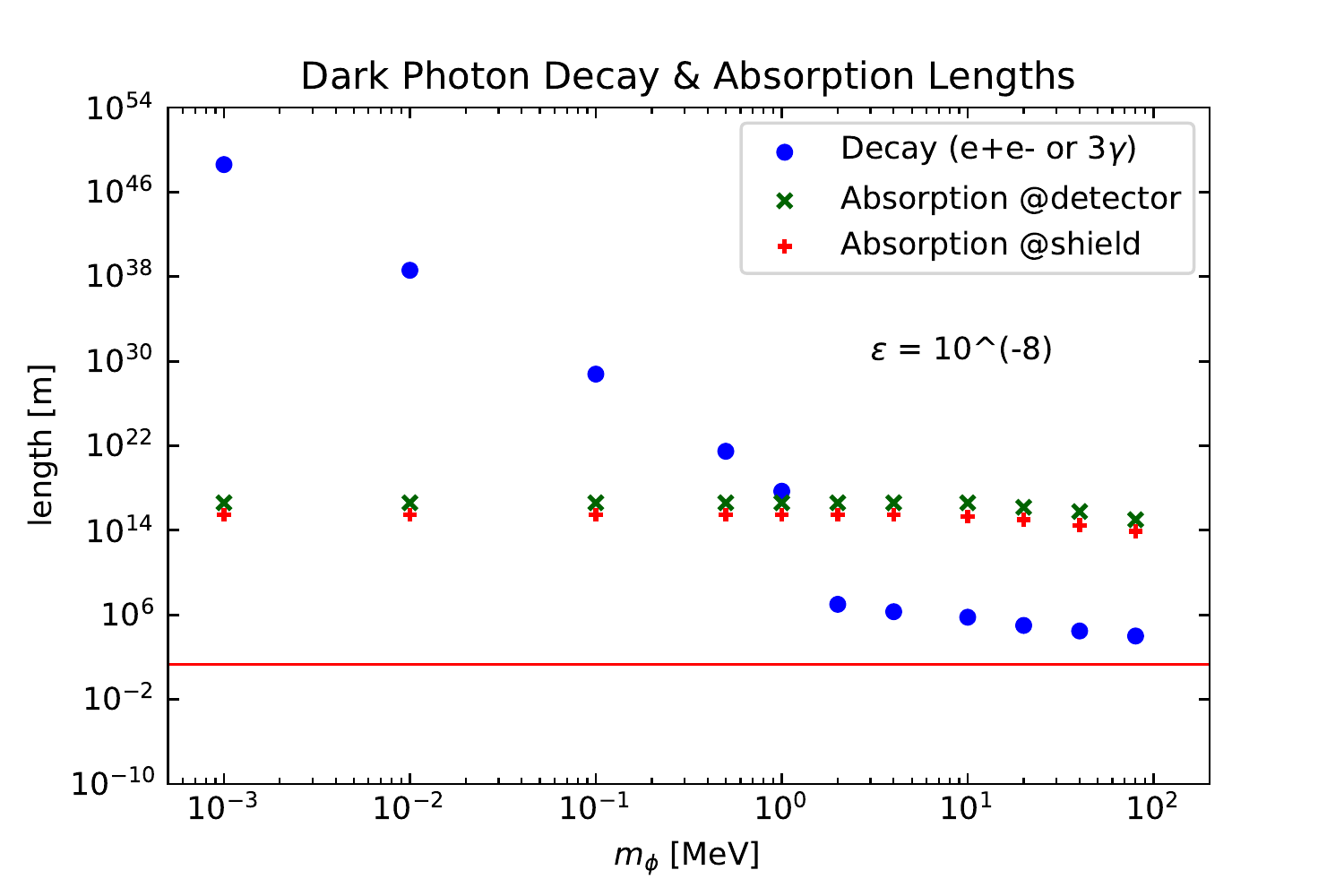}
\captionof{figure}{\label{f:dp_lengths}
  Dark photon decay and absorption lengths in the Yemilab setup for a 100~MeV-100~kW $e^-$ beam
  incident on a tungsten target (50~cm).  The horizontal red line represents the diameter of the
  neutrino detector that is assumed in this study.
}
\end{figure}

Figure~\ref{f:dp_lengths} shows DP decay and absorption lengths for $\epsilon = 10^{-3}, 10^{-5}$ and
$10^{-8}$ cases  for several different DP masses. The red horizontal line indicates detector diameter (20~m)
where DP can either decay or be absorbed. 
If the DP decay and absorption lengths are longer or shorter than the detector size, 
the detection probability is suppressed. 

\subsection{Background estimation}
Our strategy of DP search is to find excess of events by subtracting beam-OFF data from beam-ON data. 
In this way all background can be removed, 
but the background subtraction introduces statistical uncertainty in the final number of signal events, affecting final sensitivity. 
A rough estimation of background is obtained to know its impact on DP sensitivity. 
Possible background sources are radiogenic, cosmic muon, neutron, solar $\nu$, atmospheric $\nu$,
and beam related backgrounds.  

All radiogenic background can be removed by requiring 5~MeV energy threshold. 
Cosmic muon backgrounds can be suppressed by using a pulsed beam and adding an additional muon veto system that
might be, for example, plastic scintillator modules on top of the detector plus an outer water Cherenkov veto detector
that surrounds the inner detector~\cite{Seo:2019dpr}.
Cosmogenic neutron backgrounds can be suppressed by tagging neutrons in a Gadolinium-loaded water, WbLS or LS detector.
Neutrinos from $^{8}$B decays in the solar interior can contribute to the above 5~MeV background
but most of these could be removed by a directional veto
and/or by requiring very forward vertex positions for DP signals in a water or WbLS detector. 
For an LS detector it would be challenging to apply directional veto. 
Rough estimation of the possible backgrounds assuming a LS detector is described below. 
\\

\noindent
{\bf \underline{Cosmic muon, neutron, and $^8$B solar $\nu$ background}:  }
Recently Borexino has measured $^8$B solar neutrino flux more precisely in Ref.~\cite{Agostini:2017cav}. 
In Table III of Ref.~\cite{Agostini:2017cav}, 
the $^8$B solar $\nu$ rate is listed as ``bulk event'' which also includes residual background originating from cosmic muon,
and background from neutron capture-induced gamma is listed as ``external component''. 
These background rates are listed for low energy (HER-I) and high energy (HER-II) regions defined in Ref.~\cite{Agostini:2017cav}.
Thanks to the 5~MeV energy threshold, the low energy background rate can be reduced to $\sim$$1/3$ according to Fig.2 of Ref.~\cite{Agostini:2017cav},
and $^{208}$Tl-related background at 2.6~MeV is also removed.
By combining HER-I and HER-II rates for the ``bulk event'' and ``external component'', 
we obtain 0.256 cpd/100~ton, or equivalently 935 events per year 
for 1~kton fiducial volume~\footnote{Fiducial volume should be optimized for physics analysis later.} for the Yemilab neutrino detector. 
\\

\noindent
{\bf \underline{Atmospheric $\nu$ background}: }
Atmospheric neutrinos can contribute to background via a single $\pi^0$ production in
$\nu + N \rightarrow \nu + N’ + \pi^0$ process.
However, our electron beam energy (100~MeV) is less than $\pi^0$ rest mass (135~MeV), and therefore the two photons from $\pi^0$ decay 
would fire more PMTs than those from DP signal produced by our $e^-$ beam. Thus, this background would be removed 
by requiring maximum number of fired PMTs or by requiring maximum visible energy. 
Total number of atmospheric background is estimated by Borexino measurement in Ref.~\cite{Atroshchenko:2016bpy}.
According Table 2 in \cite{Atroshchenko:2016bpy}, total number of atmospheric background events in Borexino detector 
is estimated as 6.7 events/year/100~ton between 0.25~MeV to 100~MeV visible energy. 
This gives us 67 events/year/1~kton for the Yemilab neutrino detector as a conservative number because of the 5~MeV energy threshold. 
\\

\noindent
{\bf \underline{Beam-related background}: }
There are two sources of beam-related background: one is neutrinos and the other is spallation neutrons. 
Firstly, neutrinos produced by NC or CC interactions ($e + N \rightarrow e + N + \olsi{\nu} + \nu$ or $e + N \rightarrow \nu_{e} + N^{\prime} + X$) 
in a beam target can mimic DP signal by neutrino scattering in our detector ($\nu + N \rightarrow \nu + N^{\prime} + X$ or $\nu + e \rightarrow \nu + e$). 
The production cross-section of these neutrinos is found to be small, 
$\sim$$fb$/nucleon \cite{Izaguirre:2015pva} which corresponds to $\sim$10$^8$ neutrinos per year for a tungsten target assuming 100~MeV-100~kW electron beam.  
The scattering probability for a neutrino to an electron in the detector is given by $\sigma_{\nu_e} \times n_D \times l_D$ where $\sigma_{\nu_e}$ 
is the cross-section of neutrino-electron scattering, and $n_D$ and $l_D$ are the number density for water and horizontal length (20~m) of the detector. 
The scattering cross-section sigma scales as $G_F^2 \times E_{\nu}\times m_e$ where $G_F = 1.66\times 10^{-5}/$GeV$^2$. 
Even though a maximum neutrino energy of 100~MeV is considered, the scattering probability is very small,
{\it i.e.} $\sim$10$^{-18}$ scattering per neutrino resulting in $\sim$10$^{-10}$ scatterings per year, so that this background is negligible. 
Secondly, spallation neutrons produced by inelastic scattering of electrons on nuclei target can mimic DP signal 
by the neutron scattering or neutrino (produced by beta decay of the neutrons) scattering in our detector. 
The inelastic cross-section of electrons on nuclei target can reach to $\sim$$mb$ level \cite{Izaguirre:2015pva} 
resulting in a significant number of spallation neutrons and their beta-decay neutrinos. 
The spallation neutrons can be removed by placing few meters of rocks between beam-target and detector. 
The energy of the neutrinos produced from beta decay of neutrons is less than 5~MeV and therefore they can be also removed 
by the 5~MeV energy threshold we employed in our study.

\subsection{Dark photon sensitivity}

The expected number of dark photon signal events are obtained from Eq.~(\ref{e:dp_num}), 
where the energy threshold of the detector ($E_{\textrm{cut}}$) is set at 5~MeV in order to
discriminate against all radiogenic backgrounds. 

\begin{figure*}[tbp]
\centering 
\includegraphics[width=.7\textwidth]{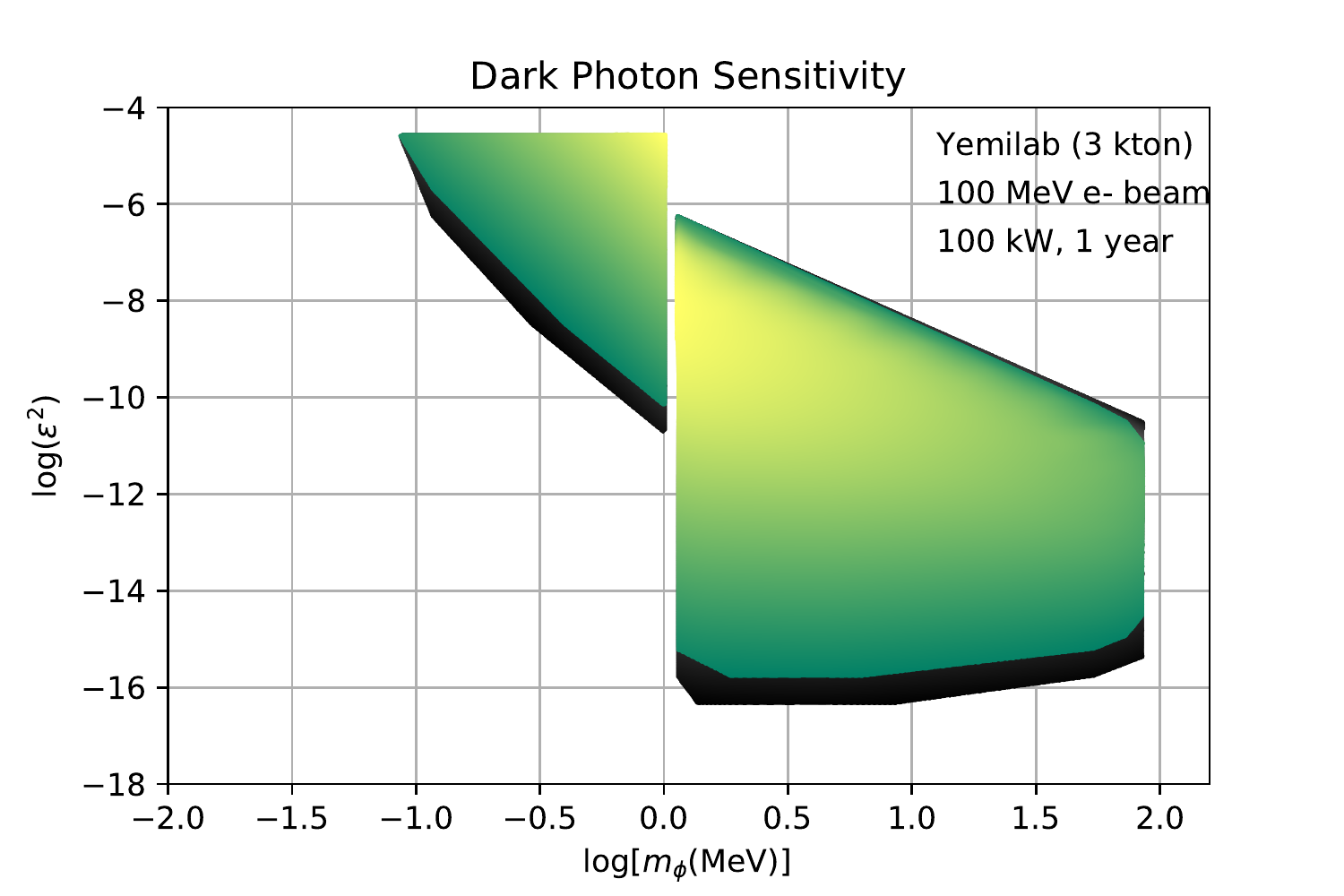}
\includegraphics[width=.7\textwidth]{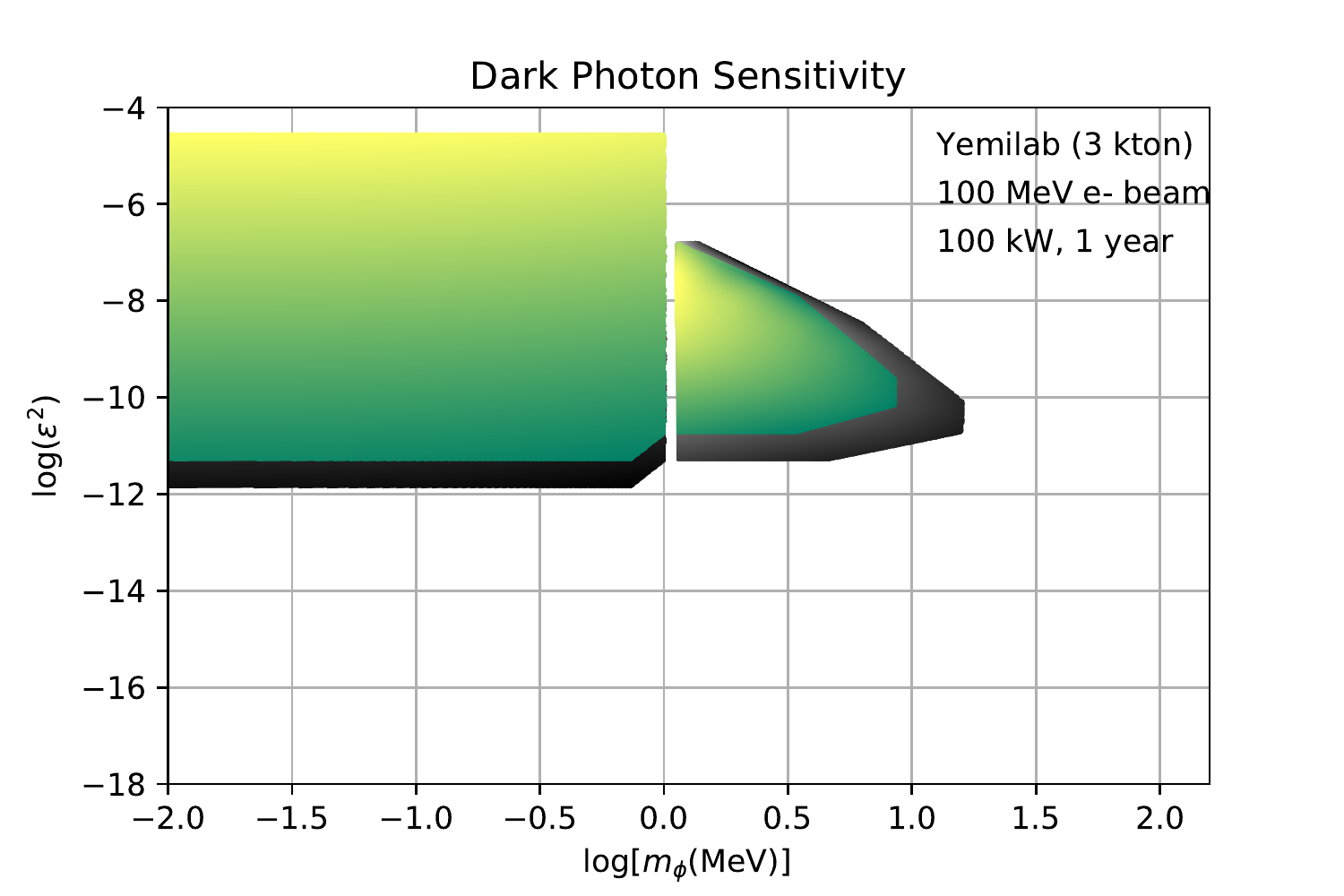}
\includegraphics[width=.7\textwidth]{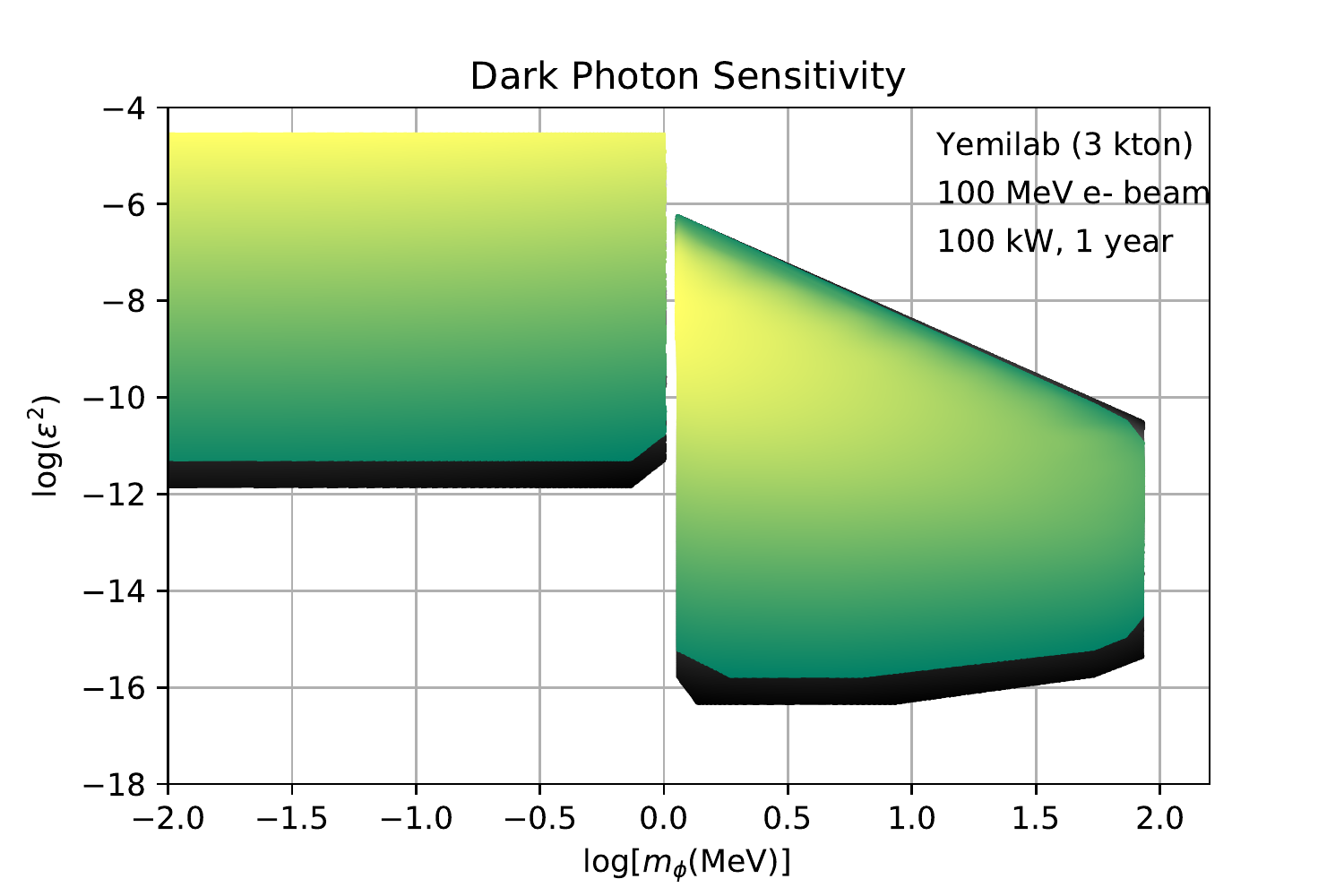}
\captionof{figure}{\label{f:dp_sens} Dark photon excluded regions (95\% C.L. sensitivities) in the Yemilab
neutrino detector for visible decays only (top),
absorption-only (middle) and for both visible decays and absorption (bottom) for 
one year of data-taking with a 100~MeV-100~kW $e^-$ beam on a tungsten target (50~cm).
Dark gray (green) region is the sensitivity obtained for zero ($10^3$) background events in both beam-ON and OFF periods.  
}
\end{figure*}

\begin{figure*}[tbp]
\centering
\includegraphics[width=1.0\textwidth]{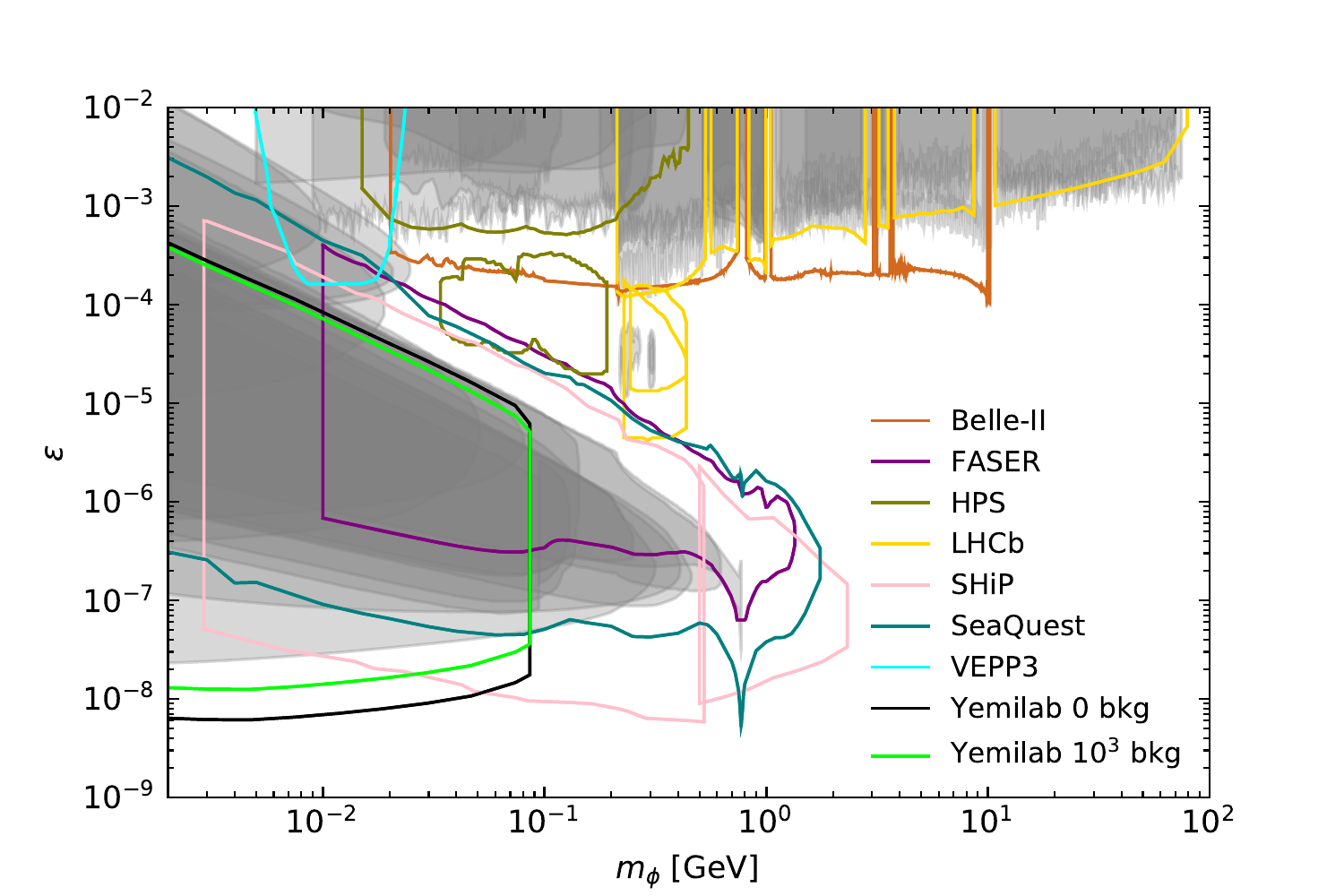}
\captionof{figure}{\label{f:dp_comp} 
Comparison of existing limits (shaded gray) and some future projections (colored lines) for DP ($m_{\phi} > 2m_e$) searches, 
drawn with darkcast framework \cite{Ilten:2018crw} at \href{https://gitlab.com/philten/darkcast}{https://gitlab.com/philten/darkcast}.
Yemilab 95\% C.L. sensitivities with zero ($10^3$) background events is shown in a black (green) line.  
Other future projections are drawn for Belle-II \cite{Kou:2018nap} (chocolate), FASER \cite{Ariga:2018uku} (purple), HPS \cite{Baltzell:2016eee} (olive),
LHCb \cite{Ilten:2016tkc} (gold), SHiP \cite{Alekhin:2015byh} (pink), SeaQuest \cite{Gardner:2015wea} (teal), and VEPP3 \cite{Wojtsekhowski:2012zq} (cyan). 
}
\end{figure*}

From the rough estimation of background discussed in the previous subsection,  
a total number of background events is expected to be $10^3$ events per year for 1~kton fiducial volume in the Yemilab neutrino detector. 
However, all of these background events can be removed by subtracting beam-OFF data from beam-ON data, 
$N_{DP} = N_{ON} - N_{OFF}\left(\frac{T_{ON}}{T_{OFF}}\right) \approx 0$. 
From the subtraction of background events, statistical fluctuation for DP signal is given as:   
\begin{equation}
\Delta N_{DP} = \sqrt{N_{ON} + \left(\frac{T_{ON}}{T_{OFF}}\right)^2 N_{OFF}} = \sqrt{N_{ON}\left(1 + \frac{T_{ON}}{T_{OFF}}\right)}
\label{e:dNDP}
\end{equation}
By taking long period of beam-OFF data, $\Delta N_{DP} \approx \sqrt{N_{ON}} \approx 32$. 
From this, one-sided 95\% C.L. sensitivity is obtained as $1.645\Delta N_{DP} \approx 53$.
Figure~\ref{f:dp_sens} shows excluded region sensitivities with 95\% C.L. assuming either zero background (dark gray region under green) 
or $10^3$ background events per year (green region) in both beam-ON and OFF data with one year operation of 100~MeV-100~kW electron beam. 
Depending on how well those estimated background to be further removed in the future, 
the 95\% C.L. exclusion sensitivity shown in green region can be improved toward dark gray region.  
The upper plot in Fig.~\ref{f:dp_sens} shows the DP sensitivity for visible
decays ($3\gamma$ or $e^+e^-$) only, where the sensitivity for
$m_\phi < 2m_e$ decreases quickly for smaller $\epsilon$ values because the $3\gamma$ decay length
exceeds the size of the detector as shown in Fig.~\ref{f:dp_lengths}; there is no sensitivity
for $\epsilon^2$ values between $10^{-6}$ and $10^{-10}$ for $m_\phi > 2m_e$ because of the short
$e^+e^-$ decay length, and below $\epsilon^2 \sim$10$^{-17}$ because of long $e^+e^-$ decay length. 
The middle plot of Fig.~\ref{f:dp_sens} shows the DP sensitivity for the absorption-only
process. 
Not shown in Fig.~\ref{f:dp_sens} is that the number of $m_{\phi}>2m_e$ DP events that are detectable via
the absorption process, which is found to be much smaller than that of those detected via the decay process
in the overlapping region of sensitivity in the parameter space.  Note also that the sensitivity
for the absorption process is nearly independent of DP mass for masses lighter than $2m_e$. 
The lack of sensitivity for sub-MeV DPs for $\epsilon^2$ values below $1.5 \times 10^{-12}$ is because
of their large absorption length.  
The bottom plot of Fig.~\ref{f:dp_sens} shows the DP sensitivity including both decay and
absorption processes for a year-long run with a 100~MeV-100~kW e$^-$ beam. 
The 95\%-C.L. exclusion level for the $e^+e^-$ decay mode for $m_{\phi}$ above $2m_e$ is
$\epsilon^2 > 4.8 \times 10^{-17}$ for zero background events, which is
comparable to that for Super-Kamiokande's zero background assumption (see Fig. 4 in Ref.~\cite{Izaguirre:2015pva}), 
and would have the world's best direct DP search sensitivity for $2m_e < m_\phi < \sim$86~MeV without considering SHiP. 
Once SHiP takes data up to $10^{20}$ POT (protons on target) \cite{Alekhin:2015byh}, 
Yemilab sensitivity is best up to $\sim$50~MeV ($\sim$20~MeV) for zero ($10^3$) background events.  
Figure~\ref{f:dp_comp} shows comparison of existing limits and some future projections of DP search for $m_{\phi} > 2m_e$. 
For sub-MeV DPs, the exclusion level is $\epsilon^2 > 1.5 \times 10^{-12}$ when considering only absorption process. 
Comparison of sub-MeV DP search is discussed in the following subsection with improved sensitivity by considering oscillation process. 

\subsection{Oscillation between ordinary and dark photons}

Although it is not shown in Fig.~\ref{f:dp_sens}, the DP sensitivity for the absorption process
extends down to very low DP masses, even to the sub-eV level. 
However, as discussed in refs.~\cite{Danilov:2018bks,Demidov:2018odn}, for $m_\phi < 2m_e$,
oscillations between ordinary and dark photons (similar to neutrino oscillations) dominate.  
The oscillation probability is given in refs.~\cite{Danilov:2018bks,Redondo:2015iea} to be 

\begin{eqnarray}
P(\gamma \rightarrow A')  &=&  \epsilon^2 \times \frac{{m_\phi}^4}{(\Delta m^2)^2 + E_{\gamma}^2\Gamma^2}, \\
P(A' \rightarrow \gamma) &=& \epsilon^2 \times \frac{{m_\phi}^4}{(\Delta m^2)^2
  + E_{\gamma}^2\Gamma^2}\times \Gamma L,
\label{e:osc-prob}
\end{eqnarray}
where
$\Delta m^2 = \sqrt{(m_\phi^2 - m_\gamma^2)^2 + 2\epsilon^2 m_\phi^2(m_\phi^2 + m_\gamma^2)} \approx |m_\phi^2 - m_\gamma^2|$, 
$m_{\gamma} = \sqrt{4\pi\alpha n_e/m_e}$ is effective photon mass in matter, $E_{\gamma}$ and $\Gamma$ are ordinary
photon energy and attenuation coefficient, respectively, and $L$ is the length of the detector in the beam
direction.

In the case of the Yemilab neutrino detector, the oscillation would have to occur twice,
one at production ($\gamma \rightarrow A'$) and the other at detection ($A' \rightarrow \gamma$).
In this case, the oscillation probability is 

\begin{equation}
  P(\gamma \leftrightarrow  A^{\prime}) = \epsilon^4 \times \frac{m_\phi^8}{\left((m_{\phi}^2 - m^{\textrm{W}\,2}_{\gamma})^2
    + E_{\gamma}^2 \Gamma_{\textrm{W}}^2 \right) \times \left((m_{\phi}^2 - m^{\textrm{H2O}\,2}_{\gamma})^2
    + E_{\gamma}^2\Gamma_{\textrm{H2O}}^2 \right)}\times \Gamma_{\textrm{H2O}} L, 
\label{e:osc-prob-yemi}
\end{equation}
where, $m^{\textrm{W}}_{\gamma}$ ($m^{\textrm{H2O}}_{\gamma}$) is an effective photon mass in tungsten (water),
i.e. 80~eV (21~eV), and $\Gamma_{\textrm{W}}$ ($\Gamma_{\textrm{H2O}}$) is a photon attenuation coefficient in
tungsten (water), where $\Gamma^{-1}_{\textrm{W}} \simeq 1$~cm ($\Gamma^{-1}_{\textrm{H2O}} \simeq 45$~cm)
at $E_{\gamma} =$ 10~MeV according to NIST database. 
In the following extreme cases, the oscillation probability Eq.~(\ref{e:osc-prob-yemi}) becomes 
\begin{eqnarray}
  P(\gamma \leftrightarrow  A^{\prime})
  &=& \epsilon^4 \times \frac{m_{\phi}^8}{m^{\textrm{W}\,4}_{\gamma} \times m^{\textrm{H2O}\,4}_{\gamma}}
                                          \times \Gamma_{\textrm{H2O}} L, \hspace{4.3cm} (m_{\phi} \ll m_{\gamma}) \\
  &=& \epsilon^4 \times \Gamma_{\textrm{H2O}} L, \hspace{7.3cm}  (m_{\phi} \gg  m_{\gamma}) \\
  &=& \epsilon^4 \times \frac{m_{\phi}^8}{ {E^2_{\gamma} \Gamma^2_{\textrm{W}}} \times 
                                            \left( (m^{\textrm{W}\,2}_{\gamma} - m^{\textrm{H2O}\,2}_{\gamma})^2
                                            + E^2_{\gamma}\Gamma^2_{\textrm{H2O}}\right)}\times \Gamma_{\textrm{H2O}} L, 
    \hspace{.1cm} (m_{\phi} \approx  m_{\gamma}^{\textrm{W}})\\
    &=& \epsilon^4 \times \frac{m_{\phi}^8} {\left( (m^{\textrm{H2O}\,2}_{\gamma} - m^{\textrm{W}\,2}_{\gamma})^2
      + E_{\gamma}^2\Gamma_{\textrm{W}}^2 \right) 
      \times E_{\gamma}^2 \Gamma_{\textrm{H2O}}^2}\times \Gamma_{\textrm{H2O}} L, \hspace{0.0cm}
    (m_{\phi} \approx  m_{\gamma}^{\textrm{H2O}})
\label{e:osc-prob-yemi-ext}
\end{eqnarray}
For $m_{\gamma}^{\textrm{H2O}} < m_{\phi} < m_{\gamma}^{\textrm{W}}$, the oscillation probability is the same as
Eq.(\ref{e:osc-prob-yemi}). 
At resonance ($m_{\phi} \approx m_{\gamma}^{\textrm{W}}$ or $m_{\gamma}^{\textrm{H2O}}$), the oscillation probability
becomes maximum. 
Using Eq.(\ref{e:osc-prob-yemi}), the expected number of DP signal events from the oscillation is obtained as 
\begin{equation}
  N_{\phi}^{\textrm{osc}} \approx N_{\textrm{e}} \times \int_{E_{\gamma}^{\textrm{min}}}^{E_{\gamma}^{\textrm{max}}}
 dE_{\gamma} P(\gamma \leftrightarrow  A^{\prime})\int_{0}^{T}dt\left(I_{\gamma}^{(1)}(t,E_{\gamma})+I_{\gamma}^{(2)}(t,E_{\gamma}) \right),
\label{e:nevt_osc-yemi}
\end{equation}
where $I^{(1)}_{\gamma}$ and $I^{(2)}_{\gamma}$ are, respectively, the 1$^{\textrm{st}}$ and 2$^{\textrm{nd}}$ generations of photon flux in target
per an incoming electron and given in the Eqs. (24) and (29) of Ref. \cite{Tsai:1966js}; 
$E_{\gamma}^{\textrm{min}} = 5$~MeV to remove radiogenic background and
$E_{\gamma}^{\textrm{max}} \approx E_{0} =$ 100~MeV.  
 
\begin{figure*}[tbp]
\centering 
\includegraphics[width=.9\textwidth]{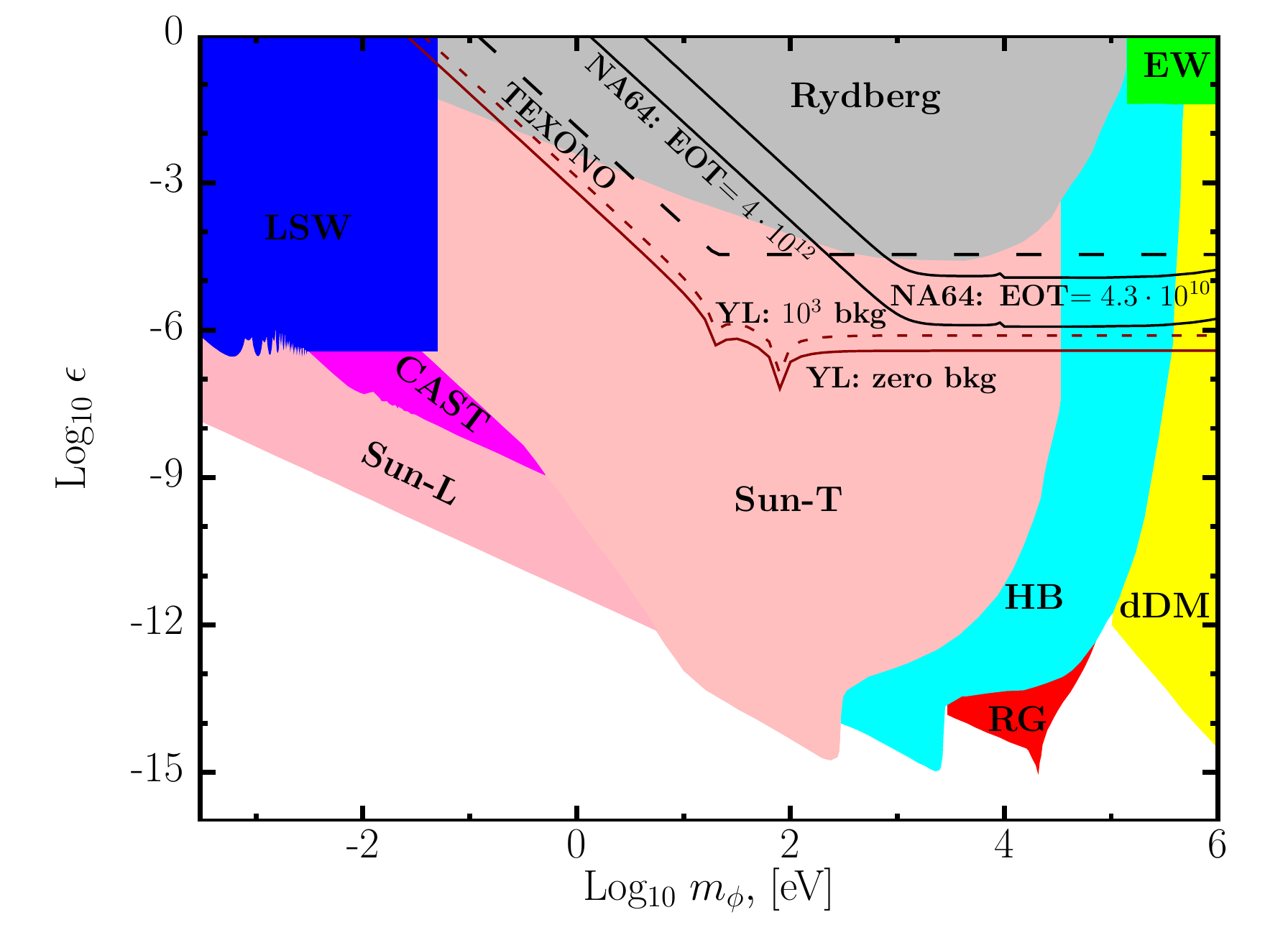}
\captionof{figure}{\label{f:dp_sens_osc}
  The dark photon sensitivity from the $\gamma \leftrightarrow A^{\prime}$ oscillation for $m_\phi < 2m_e$
  at the Yemilab (YL) neutrino detector for one year of data taking with a 100~MeV-100~kW $e^-$ beam (zero background: solid red line,
  $10^3$ background: dotted red line) on a tungsten target (50~cm), compared to those of the recent direct search experiments, 
  TEXONO (dashed black line)~\cite{Danilov:2018bks} and NA64 (two solid black lines)~\cite{Demidov:2018odn}.
  Details on the limits from the helioscopic/astrophysical observations and the other experiments are found 
  in~\cite{Redondo:2013lna,Hewett:2012ns,An:2014twa}. 
}
\end{figure*}

The 95\% C.L. sub-MeV DP detection sensitivities for photon-DP oscillations determined from Eq.~(\ref{e:nevt_osc-yemi}) 
are shown in Fig.~\ref{f:dp_sens_osc} for zero and $10^3$ background events with one year operation of 100~MeV-100~kW $e^-$ beam. 
A comparison of Fig.~\ref{f:dp_sens_osc} with Figs.~\ref{f:dp_sens} shows that 
the above 80~eV mass DP sensitivity from oscillations is better than that for the absorption process;  
the best 95\% C.L. direct DP search sensitivity, $\epsilon^2 > 1.5 \times 10^{-13} (6.1 \times 10^{-13})$ for zero ($10^3$) background events, 
is obtained in a year-long data-taking run with a 100~MeV-100~kW e$^{-}$ beam on a thick tungsten target
and the Yemilab neutrino detector.

\section{Summary}
\label{sec:sum}

Dark photon searches are the focus of a variety of experiments and have been invoked to explain
a number of anomalies that have cropped up in (astro-)particle physics observations.  Many of the
best constraints, especially for sub-MeV dark photons, are from helioscopic or astrophysical observations. 
However, the helioscopic/astrophysical constraints depend, in a large part, on the choice of DP
mean-free-path lengths inside stellar objects~\cite{Redondo:2008aa} and, therefore, direct search
experiments at laboratories are absolutely necessary. 
Our study shows that a combination of a 3~kiloton-scale neutrino detector and an electron beam at
Yemilab could constrain DP kinetic mixing parameters with the world's best direct search sensitivity
for sub-MeV and above MeV DPs produced via darkstrahlung ($e^- + Z \rightarrow e^- + Z + A^{\prime}$)
or oscillations ($\gamma \rightarrow A^{\prime}$).  
In our study, DP is searched by observing excess of events in beam-ON data subtracted by beam-OFF data. 
By detecting DPs via their absorption ($A^\prime + e^- \rightarrow \gamma + e^-$) or oscillations
($A^{\prime} \rightarrow \gamma$) processes, a 95\% C.L. direct search sensitivity for 80~eV $< m_{\phi} < 2m_e$ MeV
DP of $\epsilon^2 > 1.5 \times 10^{-13} (6.1 \times 10^{-13})$ could be achieved with one year of
operation of a 100~MeV-100~kW electron beam on a thick tungsten target for zero ($10^3$) background events in both beam-ON and OFF periods.  
The best sensitivities for sub-MeV DPs are achieved by exploiting the oscillation between ordinary and dark photons. 
At the peaks of the oscillation resonances, i.e. $m_{\phi} =$ 21~eV (in water) and 80~eV (in tungsten),
the sensitivities are enhanced as shown in Fig.~\ref{f:dp_sens_osc}, 
and $\epsilon^2 \sim\mathcal{O}(10^{-15})$ at $m_{\phi} =$ 80~eV for zero background events. 
The dark photon sensitivity for masses below the $m_{\phi} <$ 21~eV resonance peak rapidly decreases
as $\epsilon^4 \propto m_\phi^{-8}$ due to the oscillation. 
Sub-MeV dark photon detection via decays to 3$\gamma$ are highly suppressed because of the very long decay lengths. 
For $2m_e < m_\phi < \sim$86~MeV DPs, the direct search sensitivity for the kinetic mixing
parameters using visible decays ($A^\prime \rightarrow e^+e^-$) is $\epsilon^2 > \mathcal{O}(10^{-17}) (\mathcal{O}(10^{-16}))$
at the 95\% C.L. for zero ($10^3$) background events using the 100~MeV-100~kW $e^-$ beam; 
a sensitivity that is comparable to that of Super-K for the same zero background assumption as Super-K. 
With a higher energy electron beam, the sensitivity beyond 86~MeV DPs could also be explored but, 
it may not be practical to accommodate such a facility in the current Yemilab design configuration.



\acknowledgments
The authors are grateful to D. Gorbunov and S. Demidov for very useful discussions about
the possibilities of oscillations between ordinary and dark photons
as well as providing us with the data points incorporated in Fig.~\ref{f:dp_sens_osc}. 
We also thank H.S. Lee and H.K. Park for visiting IBS to share their expertise on dark photons
and thank T. Terada for help with the Mathematica calculations. 
Our special thanks goes to S. Olsen for reading our manuscript carefully and for giving us valuable feedback. 
This work was supported by the National Research Foundation of Korea (NRF) grant funded by the Korea Ministry of
Science and ICT (MSIT) (No. 2017R1A2B4012757, IBS-R016-D1-2019-b01) and IBS-R016-D1. 

\bibliographystyle{JHEP}
\bibliography{references}

\providecommand{\href}[2]{#2}\begingroup\raggedright\begin{thebibliography}{10}

\bibitem{Krasznahorkay:2015iga}
A.~J. Krasznahorkay et~al., \emph{{Observation of Anomalous Internal Pair
  Creation in Be8 : A Possible Indication of a Light, Neutral Boson}},
  \href{https://doi.org/10.1103/PhysRevLett.116.042501}{\emph{Phys. Rev. Lett.}
  {\bfseries 116} (2016) 042501}
  [\href{https://arxiv.org/abs/1504.01527}{{\ttfamily 1504.01527}}].

\bibitem{Krasznahorkay:2019lyl}
A.~Krasznahorkay et~al., \emph{{New evidence supporting the existence of the
  hypothetic X17 particle}},
  \href{https://arxiv.org/abs/1910.10459}{{\ttfamily 1910.10459}}.

\bibitem{Bjorken:2009mm}
J.~D. Bjorken, R.~Essig, P.~Schuster and N.~Toro, \emph{{New Fixed-Target
  Experiments to Search for Dark Gauge Forces}},
  \href{https://doi.org/10.1103/PhysRevD.80.075018}{\emph{Phys. Rev. D}
  {\bfseries 80} (2009) 075018}
  [\href{https://arxiv.org/abs/0906.0580}{{\ttfamily 0906.0580}}].

\bibitem{Bjorken:1988as}
J.~Bjorken, S.~Ecklund, W.~Nelson, A.~Abashian, C.~Church, B.~Lu et~al.,
  \emph{{Search for Neutral Metastable Penetrating Particles Produced in the
  SLAC Beam Dump}}, \href{https://doi.org/10.1103/PhysRevD.38.3375}{\emph{Phys.
  Rev. D} {\bfseries 38} (1988) 3375}.

\bibitem{Bross:1989mp}
A.~Bross, M.~Crisler, S.~H. Pordes, J.~Volk, S.~Errede and J.~Wrbanek, \emph{{A
  Search for Shortlived Particles Produced in an Electron Beam Dump}},
  \href{https://doi.org/10.1103/PhysRevLett.67.2942}{\emph{Phys. Rev. Lett.}
  {\bfseries 67} (1991) 2942}.

\bibitem{Merkel:2011ze}
{\scshape A1} collaboration, \emph{{Search for Light Gauge Bosons of the Dark
  Sector at the Mainz Microtron}},
  \href{https://doi.org/10.1103/PhysRevLett.106.251802}{\emph{Phys. Rev. Lett.}
  {\bfseries 106} (2011) 251802}
  [\href{https://arxiv.org/abs/1101.4091}{{\ttfamily 1101.4091}}].

\bibitem{Abrahamyan:2011gv}
{\scshape APEX} collaboration, \emph{{Search for a New Gauge Boson in
  Electron-Nucleus Fixed-Target Scattering by the APEX Experiment}},
  \href{https://doi.org/10.1103/PhysRevLett.107.191804}{\emph{Phys. Rev. Lett.}
  {\bfseries 107} (2011) 191804}
  [\href{https://arxiv.org/abs/1108.2750}{{\ttfamily 1108.2750}}].

\bibitem{Essig:2010xa}
R.~Essig, P.~Schuster, N.~Toro and B.~Wojtsekhowski, \emph{{An Electron Fixed
  Target Experiment to Search for a New Vector Boson A' Decaying to e+e-}},
  \href{https://doi.org/10.1007/JHEP02(2011)009}{\emph{JHEP} {\bfseries 02}
  (2011) 009} [\href{https://arxiv.org/abs/1001.2557}{{\ttfamily 1001.2557}}].

\bibitem{Babusci:2012cr}
{\scshape KLOE-2} collaboration, \emph{{Limit on the production of a light
  vector gauge boson in phi meson decays with the KLOE detector}},
  \href{https://doi.org/10.1016/j.physletb.2013.01.067}{\emph{Phys. Lett. B}
  {\bfseries 720} (2013) 111}
  [\href{https://arxiv.org/abs/1210.3927}{{\ttfamily 1210.3927}}].

\bibitem{Izaguirre:2013uxa}
E.~Izaguirre, G.~Krnjaic, P.~Schuster and N.~Toro, \emph{{New Electron
  Beam-Dump Experiments to Search for MeV to few-GeV Dark Matter}},
  \href{https://doi.org/10.1103/PhysRevD.88.114015}{\emph{Phys. Rev. D}
  {\bfseries 88} (2013) 114015}
  [\href{https://arxiv.org/abs/1307.6554}{{\ttfamily 1307.6554}}].

\bibitem{Babusci:2014sta}
{\scshape KLOE-2} collaboration, \emph{{Search for light vector boson
  production in $e^+e^- \rightarrow \mu^+ \mu^- \gamma$ interactions with the
  KLOE experiment}},
  \href{https://doi.org/10.1016/j.physletb.2014.08.005}{\emph{Phys. Lett. B}
  {\bfseries 736} (2014) 459}
  [\href{https://arxiv.org/abs/1404.7772}{{\ttfamily 1404.7772}}].

\bibitem{Batell:2014mga}
B.~Batell, R.~Essig and Z.~Surujon, \emph{{Strong Constraints on Sub-GeV Dark
  Sectors from SLAC Beam Dump E137}},
  \href{https://doi.org/10.1103/PhysRevLett.113.171802}{\emph{Phys. Rev. Lett.}
  {\bfseries 113} (2014) 171802}
  [\href{https://arxiv.org/abs/1406.2698}{{\ttfamily 1406.2698}}].

\bibitem{Merkel:2014avp}
H.~Merkel et~al., \emph{{Search at the Mainz Microtron for Light Massive Gauge
  Bosons Relevant for the Muon g-2 Anomaly}},
  \href{https://doi.org/10.1103/PhysRevLett.112.221802}{\emph{Phys. Rev. Lett.}
  {\bfseries 112} (2014) 221802}
  [\href{https://arxiv.org/abs/1404.5502}{{\ttfamily 1404.5502}}].

\bibitem{Blumlein:2013cua}
J.~Bl{\"u}mlein and J.~Brunner, \emph{{New Exclusion Limits on Dark Gauge
  Forces from Proton Bremsstrahlung in Beam-Dump Data}},
  \href{https://doi.org/10.1016/j.physletb.2014.02.029}{\emph{Phys. Lett. B}
  {\bfseries 731} (2014) 320}
  [\href{https://arxiv.org/abs/1311.3870}{{\ttfamily 1311.3870}}].

\bibitem{Batley:2015lha}
{\scshape NA48/2} collaboration, \emph{{Search for the dark photon in $\pi^0$
  decays}}, \href{https://doi.org/10.1016/j.physletb.2015.04.068}{\emph{Phys.
  Lett. B} {\bfseries 746} (2015) 178}
  [\href{https://arxiv.org/abs/1504.00607}{{\ttfamily 1504.00607}}].

\bibitem{Anastasi:2015qla}
A.~Anastasi et~al., \emph{{Limit on the production of a low-mass vector boson
  in $\mathrm{e}^{+}\mathrm{e}^{-} \to \mathrm{U}\gamma$, $\mathrm{U} \to
  \mathrm{e}^{+}\mathrm{e}^{-}$ with the KLOE experiment}},
  \href{https://doi.org/10.1016/j.physletb.2015.10.003}{\emph{Phys. Lett. B}
  {\bfseries 750} (2015) 633}
  [\href{https://arxiv.org/abs/1509.00740}{{\ttfamily 1509.00740}}].

\bibitem{Ilten:2015hya}
P.~Ilten, J.~Thaler, M.~Williams and W.~Xue, \emph{{Dark photons from charm
  mesons at LHCb}},
  \href{https://doi.org/10.1103/PhysRevD.92.115017}{\emph{Phys. Rev. D}
  {\bfseries 92} (2015) 115017}
  [\href{https://arxiv.org/abs/1509.06765}{{\ttfamily 1509.06765}}].

\bibitem{Anastasi:2016ktq}
{\scshape KLOE-2} collaboration, \emph{{Limit on the production of a new vector
  boson in $\mathrm{e^+ e^-}\rightarrow {\rm U}\gamma$, U$\rightarrow
  \pi^+\pi^-$ with the KLOE experiment}},
  \href{https://doi.org/10.1016/j.physletb.2016.04.019}{\emph{Phys. Lett. B}
  {\bfseries 757} (2016) 356}
  [\href{https://arxiv.org/abs/1603.06086}{{\ttfamily 1603.06086}}].

\bibitem{Ilten:2016tkc}
P.~Ilten, Y.~Soreq, J.~Thaler, M.~Williams and W.~Xue, \emph{{Proposed
  Inclusive Dark Photon Search at LHCb}},
  \href{https://doi.org/10.1103/PhysRevLett.116.251803}{\emph{Phys. Rev. Lett.}
  {\bfseries 116} (2016) 251803}
  [\href{https://arxiv.org/abs/1603.08926}{{\ttfamily 1603.08926}}].

\bibitem{Battaglieri:2017qen}
{\scshape BDX} collaboration, \emph{{Dark matter search in a Beam-Dump
  eXperiment (BDX) at Jefferson Lab: an update on PR12-16-001}},
  \href{https://arxiv.org/abs/1712.01518}{{\ttfamily 1712.01518}}.

\bibitem{Lees:2017lec}
{\scshape BaBar} collaboration, \emph{{Search for Invisible Decays of a Dark
  Photon Produced in ${e}^{+}{e}^{-}$ Collisions at BaBar}},
  \href{https://doi.org/10.1103/PhysRevLett.119.131804}{\emph{Phys. Rev. Lett.}
  {\bfseries 119} (2017) 131804}
  [\href{https://arxiv.org/abs/1702.03327}{{\ttfamily 1702.03327}}].

\bibitem{Corliss:2017tms}
{\scshape DarkLight} collaboration, \emph{{Searching for a dark photon with
  DarkLight}}, \href{https://doi.org/10.1016/j.nima.2016.07.053}{\emph{Nucl.
  Instrum. Meth. A} {\bfseries 865} (2017) 125}.

\bibitem{Aaij:2017rft}
{\scshape LHCb} collaboration, \emph{{Search for Dark Photons Produced in 13
  TeV $pp$ Collisions}},
  \href{https://doi.org/10.1103/PhysRevLett.120.061801}{\emph{Phys. Rev. Lett.}
  {\bfseries 120} (2018) 061801}
  [\href{https://arxiv.org/abs/1710.02867}{{\ttfamily 1710.02867}}].

\bibitem{Ilten:2018crw}
P.~Ilten, Y.~Soreq, M.~Williams and W.~Xue, \emph{{Serendipity in dark photon
  searches}}, \href{https://doi.org/10.1007/JHEP06(2018)004}{\emph{JHEP}
  {\bfseries 06} (2018) 004}
  [\href{https://arxiv.org/abs/1801.04847}{{\ttfamily 1801.04847}}].

\bibitem{Ariga:2018uku}
{\scshape FASER} collaboration, \emph{{FASER\textquoteright{}s physics reach
  for long-lived particles}},
  \href{https://doi.org/10.1103/PhysRevD.99.095011}{\emph{Phys. Rev. D}
  {\bfseries 99} (2019) 095011}
  [\href{https://arxiv.org/abs/1811.12522}{{\ttfamily 1811.12522}}].

\bibitem{Banerjee:2019hmi}
{\scshape NA64} collaboration, \emph{{Improved limits on a hypothetical X(16.7)
  boson and a dark photon decaying into $e^+e^-$ pairs}},
  \href{https://doi.org/10.1103/PhysRevD.101.071101}{\emph{Phys. Rev. D}
  {\bfseries 101} (2020) 071101}
  [\href{https://arxiv.org/abs/1912.11389}{{\ttfamily 1912.11389}}].

\bibitem{Demidov:2018odn}
S.~Demidov, S.~Gninenko and D.~Gorbunov, \emph{{Light hidden photon production
  in high energy collisions}},
  \href{https://doi.org/10.1007/JHEP07(2019)162}{\emph{JHEP} {\bfseries 07}
  (2019) 162} [\href{https://arxiv.org/abs/1812.02719}{{\ttfamily
  1812.02719}}].

\bibitem{Park:2017prx}
H.~Park, \emph{{Detecting Dark Photons with Reactor Neutrino Experiments}},
  \href{https://doi.org/10.1103/PhysRevLett.119.081801}{\emph{Phys. Rev. Lett.}
  {\bfseries 119} (2017) 081801}
  [\href{https://arxiv.org/abs/1705.02470}{{\ttfamily 1705.02470}}].

\bibitem{Danilov:2018bks}
M.~Danilov, S.~Demidov and D.~Gorbunov, \emph{{Constraints on hidden photons
  produced in nuclear reactors}},
  \href{https://doi.org/10.1103/PhysRevLett.122.041801}{\emph{Phys. Rev. Lett.}
  {\bfseries 122} (2019) 041801}
  [\href{https://arxiv.org/abs/1804.10777}{{\ttfamily 1804.10777}}].

\bibitem{Beacham:2019nyx}
J.~Beacham et~al., \emph{{Physics Beyond Colliders at CERN: Beyond the Standard
  Model Working Group Report}},
  \href{https://doi.org/10.1088/1361-6471/ab4cd2}{\emph{J. Phys. G} {\bfseries
  47} (2020) 010501} [\href{https://arxiv.org/abs/1901.09966}{{\ttfamily
  1901.09966}}].

\bibitem{Filippi:2020kii}
A.~Filippi and M.~De~Napoli, \emph{{Searching in the dark: the hunt for the
  dark photon}}, \href{https://doi.org/10.1016/j.revip.2020.100042}{\emph{Rev.
  Phys.} {\bfseries 5} (2020) 100042}
  [\href{https://arxiv.org/abs/2006.04640}{{\ttfamily 2006.04640}}].

\bibitem{Essig:2013lka}
R.~Essig et~al., \emph{{Working Group Report: New Light Weakly Coupled
  Particles}},  in \emph{{Community Summer Study 2013}: {Snowmass on the
  Mississippi}}, 10, 2013, \href{https://arxiv.org/abs/1311.0029}{{\ttfamily
  1311.0029}}.

\bibitem{Aaij:2019bvg}
{\scshape LHCb} collaboration, \emph{{Search for $A'\to\mu^+\mu^-$ Decays}},
  \href{https://doi.org/10.1103/PhysRevLett.124.041801}{\emph{Phys. Rev. Lett.}
  {\bfseries 124} (2020) 041801}
  [\href{https://arxiv.org/abs/1910.06926}{{\ttfamily 1910.06926}}].

\bibitem{Adhikari:2019off}
{\scshape COSINE-100} collaboration, \emph{{Search for a Dark Matter-Induced
  Annual Modulation Signal in NaI(Tl) with the COSINE-100 Experiment}},
  \href{https://doi.org/10.1103/PhysRevLett.123.031302}{\emph{Phys. Rev. Lett.}
  {\bfseries 123} (2019) 031302}
  [\href{https://arxiv.org/abs/1903.10098}{{\ttfamily 1903.10098}}].

\bibitem{Kim:2017xrs}
G.~Kim et~al., \emph{{Novel measurement method of heat and light detection for
  neutrinoless double beta decay}},
  \href{https://doi.org/10.1016/j.astropartphys.2017.02.009}{\emph{Astropart.
  Phys.} {\bfseries 91} (2017) 105}.

\bibitem{Seo:2019dpr}
S.-H. Seo, \emph{{Neutrino Telescope at Yemilab, Korea}},
  \href{https://arxiv.org/abs/1903.05368}{{\ttfamily 1903.05368}}.

\bibitem{Seo:2016uom}
{\scshape RENO} collaboration, \emph{{Spectral Measurement of the Electron
  Antineutrino Oscillation Amplitude and Frequency using 500 Live Days of RENO
  Data}}, \href{https://doi.org/10.1103/PhysRevD.98.012002}{\emph{Phys. Rev. D}
  {\bfseries 98} (2018) 012002}
  [\href{https://arxiv.org/abs/1610.04326}{{\ttfamily 1610.04326}}].

\bibitem{Ko:2016owz}
{\scshape NEOS} collaboration, \emph{{Sterile Neutrino Search at the NEOS
  Experiment}},
  \href{https://doi.org/10.1103/PhysRevLett.118.121802}{\emph{Phys. Rev. Lett.}
  {\bfseries 118} (2017) 121802}
  [\href{https://arxiv.org/abs/1610.05134}{{\ttfamily 1610.05134}}].

\bibitem{Izaguirre:2015pva}
E.~Izaguirre, G.~Krnjaic and M.~Pospelov, \emph{{MeV-Scale Dark Matter Deep
  Underground}}, \href{https://doi.org/10.1103/PhysRevD.92.095014}{\emph{Phys.
  Rev. D} {\bfseries 92} (2015) 095014}
  [\href{https://arxiv.org/abs/1507.02681}{{\ttfamily 1507.02681}}].

\bibitem{Liu:2017htz}
Y.-S. Liu and G.~A. Miller, \emph{{Validity of the Weizs\"{a}cker-Williams
  approximation and the analysis of beam dump experiments: Production of an
  axion, a dark photon, or a new axial-vector boson}},
  \href{https://doi.org/10.1103/PhysRevD.96.016004}{\emph{Phys. Rev. D}
  {\bfseries 96} (2017) 016004}
  [\href{https://arxiv.org/abs/1705.01633}{{\ttfamily 1705.01633}}].

\bibitem{Tsai:1966js}
Y.-S. Tsai and V.~Whitis, \emph{{Thick Target Bremsstrahlung and Target
  Consideration for Secondary Particle Production by Electrons}},
  \href{https://doi.org/10.1103/PhysRev.149.1248}{\emph{Phys. Rev.} {\bfseries
  149} (1966) 1248}.

\bibitem{Agostini:2017cav}
{\scshape Borexino} collaboration, \emph{{Improved measurement of $^8$B solar
  neutrinos with $1.5 kt·y$ of Borexino exposure}},
  \href{https://doi.org/10.1103/PhysRevD.101.062001}{\emph{Phys. Rev. D}
  {\bfseries 101} (2020) 062001}
  [\href{https://arxiv.org/abs/1709.00756}{{\ttfamily 1709.00756}}].

\bibitem{Atroshchenko:2016bpy}
V.~Atroshchenko and E.~Litvinovich, \emph{{Estimation of atmospheric neutrinos
  background in Borexino}},
  \href{https://doi.org/10.1088/1742-6596/675/1/012014}{\emph{J. Phys. Conf.
  Ser.} {\bfseries 675} (2016) 012014}.

\bibitem{Alekhin:2015byh}
S.~Alekhin et~al., \emph{{A facility to Search for Hidden Particles at the CERN
  SPS: the SHiP physics case}},
  \href{https://doi.org/10.1088/0034-4885/79/12/124201}{\emph{Rept. Prog.
  Phys.} {\bfseries 79} (2016) 124201}
  [\href{https://arxiv.org/abs/1504.04855}{{\ttfamily 1504.04855}}].

\bibitem{Kou:2018nap}
E.~Kou et~al., \emph{{The Belle II Physics Book}},
  \href{https://arxiv.org/abs/1808.10567}{{\ttfamily 1808.10567}}.

\bibitem{Baltzell:2016eee}
{\scshape HPS} collaboration, \emph{{The Heavy Photon Search beamline and its
  performance}}, \href{https://doi.org/10.1016/j.nima.2017.03.061}{\emph{Nucl.
  Instrum. Meth.} {\bfseries A859} (2017) 69}
  [\href{https://arxiv.org/abs/1612.07821}{{\ttfamily 1612.07821}}].

\bibitem{Gardner:2015wea}
S.~Gardner, R.~J. Holt and A.~S. Tadepalli, \emph{{New Prospects in Fixed
  Target Searches for Dark Forces with the SeaQuest Experiment at Fermilab}},
  \href{https://doi.org/10.1103/PhysRevD.93.115015}{\emph{Phys. Rev.}
  {\bfseries D93} (2016) 115015}
  [\href{https://arxiv.org/abs/1509.00050}{{\ttfamily 1509.00050}}].

\bibitem{Wojtsekhowski:2012zq}
B.~Wojtsekhowski, D.~Nikolenko and I.~Rachek, \emph{{Searching for a new force
  at VEPP-3}},  \href{https://arxiv.org/abs/1207.5089}{{\ttfamily 1207.5089}}.

\bibitem{Redondo:2015iea}
J.~Redondo, \emph{{Atlas of solar hidden photon emission}},
  \href{https://doi.org/10.1088/1475-7516/2015/07/024}{\emph{JCAP} {\bfseries
  07} (2015) 024} [\href{https://arxiv.org/abs/1501.07292}{{\ttfamily
  1501.07292}}].

\bibitem{Redondo:2013lna}
J.~Redondo and G.~Raffelt, \emph{{Solar constraints on hidden photons
  re-visited}},
  \href{https://doi.org/10.1088/1475-7516/2013/08/034}{\emph{JCAP} {\bfseries
  08} (2013) 034} [\href{https://arxiv.org/abs/1305.2920}{{\ttfamily
  1305.2920}}].

\bibitem{Hewett:2012ns}
\emph{{Fundamental Physics at the Intensity Frontier}}, 5, 2012.
\newblock 10.2172/1042577.

\bibitem{An:2014twa}
H.~An, M.~Pospelov, J.~Pradler and A.~Ritz, \emph{{Direct Detection Constraints
  on Dark Photon Dark Matter}},
  \href{https://doi.org/10.1016/j.physletb.2015.06.018}{\emph{Phys. Lett. B}
  {\bfseries 747} (2015) 331}
  [\href{https://arxiv.org/abs/1412.8378}{{\ttfamily 1412.8378}}].

\bibitem{Redondo:2008aa}
J.~Redondo, \emph{{Helioscope Bounds on Hidden Sector Photons}},
  \href{https://doi.org/10.1088/1475-7516/2008/07/008}{\emph{JCAP} {\bfseries
  07} (2008) 008} [\href{https://arxiv.org/abs/0801.1527}{{\ttfamily
  0801.1527}}].

\end{thebibliography}\endgroup

\end{document}